\documentclass[fleqn,usenatbib]{mnras}

\usepackage{newtxtext,newtxmath}

\usepackage[T1]{fontenc}
\usepackage{ae,aecompl}
\usepackage{times}

\usepackage{graphicx}	
\usepackage{amsmath}	

\newcommand{\peryr}                    {\,{\rm yr}^{-1}}

\newcommand{\kpc}                      {\,{\rm kpc}}

\newcommand{\ckpc}                      {\,{\rm ckpc}}
\newcommand{\cMpc}                      {\,{\rm cMpc}}
\newcommand{\Msun}                    {\,{\rm M}_\odot}

\newcommand{\K}                          {\,{\rm K}}

\newcommand{\dex}              {{\rm dex}}

\newcommand{\kappaco}{$\kappa_{\rm co}$ }

\title[Galaxy simulations with controlled halo assembly]{Quenching and morphological evolution due to circumgalactic gas expulsion in a simulated galaxy with a controlled assembly history}

\author[J. J. Davies, R. A. Crain and A. Pontzen]{Jonathan  J. Davies,$^{1,2}$\thanks{E-mail: j.j.davies@ucl.ac.uk}
Robert A. Crain,$^{1}$
and Andrew Pontzen$^{2}$
\\
$^{1}$Astrophysics Research Institute, Liverpool John Moores University, 146 Brownlow Hill, Liverpool L3 5RF, UK\\
$^{2}$Department of Physics and Astronomy, University College London, Gower Street, London WC1E 6BT, UK
}

\date{Accepted XXX. Received YYY; in original form ZZZ}

\pubyear{2020}

\begin{document}
\label{firstpage}
\pagerange{\pageref{firstpage}--\pageref{lastpage}}
\maketitle
\begin{abstract}
We examine the influence of dark matter halo assembly on the evolution of a simulated $\sim L^\star$ galaxy. Starting from a zoom-in simulation of a star-forming galaxy evolved with the EAGLE galaxy formation model, we use the genetic modification technique to create a pair of complementary assembly histories: one in which the halo assembles later than in the unmodified case, and one in which it assembles earlier. Delayed assembly leads to the galaxy exhibiting a greater present-day star formation rate than its unmodified counterpart, whilst in the accelerated case the galaxy quenches at $z\simeq 1$, and becomes spheroidal. We simulate each assembly history nine times, adopting different seeds for the random number generator used by EAGLE's stochastic subgrid implementations of star formation and feedback. The systematic changes driven by differences in assembly history are significantly stronger than the random scatter induced by this stochasticity. The sensitivity of $\sim L^\star$ galaxy evolution to dark matter halo assembly follows from the close coupling of the growth histories of the central black hole (BH) and the halo, such that earlier assembly fosters the formation of a more massive BH, and more efficient expulsion of circumgalactic gas. In response to this expulsion, the circumgalactic medium reconfigures at a lower density, extending its cooling time and thus inhibiting the replenishment of the interstellar medium. Our results indicate that halo assembly history significantly influences the evolution of $\sim L^\star$ central galaxies, and that the expulsion of circumgalactic gas is a crucial step in quenching them.
\end{abstract}




\section{Introduction}
\label{sec:intro}

Galaxy surveys have revealed that the present-day galaxy population can be broadly categorised into two populations within the colour vs. stellar mass parameter space: the ``blue cloud'' of star-forming and usually disc-dominated galaxies, and the ``red sequence'' of mostly quiescent, spheroidal or elliptical galaxies \citep[e.g.][]{baldry04,balogh04,baldry06,muzzin13}. The relative scarcity of galaxies in the ``green valley'' that separates these two populations implies that the transition of galaxies from the blue cloud to the red sequence must occur rapidly in a ``quenching'' process; unveiling how this process occurs is a key aim of galaxy formation theory.

For a galaxy to transition from the blue cloud to the red sequence, it must either exhaust or otherwise lose its supply of star-forming gas, and prevent it from being efficiently replenished by cooling flows \citep[e.g.][]{schawinski14}. Various mechanisms for this have been proposed, which broadly fall into two categories that are distinct and separable in the context of galaxy surveys: ``mass quenching'' and ``environment quenching'' \citep{peng10}. Environment quenching is most relevant for satellite galaxies \citep{peng12,woo15}; example mechanisms include the removal of gas by ram-pressure stripping, or by tidal forces \citep{gunn72,vandenbosch08,mccarthy08,wetzel13,kukstas20}. Mass quenching mechanisms are purely internal and apply to all galaxies; they are so named because their operation is strongly correlated with stellar mass.

The most commonly considered internal mechanism for quenching is the injection of feedback energy, which can in principle both eject gas from the galaxy and provide heat to offset cooling flows \citep[e.g.][]{wf91}. Feedback associated with the formation and evolution of massive stars has long been implemented in galaxy formation models as a means to regulate star formation in low-mass galaxies \citep[e.g.][]{navarro93,cole00,springel03,dallavecchiaschaye08,dallavecchiaschaye12,henriques13,stinson13}, however in higher-mass systems the energy available from massive stars and supernovae is insufficient for regulating galaxy growth to observed levels \citep[e.g.][]{crain09,henriques19}. To reproduce the high-mass ends of both the galaxy stellar mass function and red sequence by quenching star formation on the mass scales of $\sim L^\star$ galaxies and above, modern models typically invoke feedback from active galactic nuclei (AGN) \citep[e.g.][]{dimatteo05,bower06,bower12,croton06,somerville08,vogelsberger14,schaye15,dubois16,mccarthy17,kaviraj17,tremmel17,henden18,weinberger18,dave19}. This mechanism is motivated by observations of AGN-driven, mass-loaded outflows at both low and high redshift \citep[e.g.][]{rupke11,maiolino12,harrison14,cicone15,cicone16,bae17,rupke17,bischetti19,fluetsch19}, and by observed correlations between quenching and both the central black hole mass \citep{terrazas16,terrazas17,martinnavarro16,martinnavarro18} and proxies for it \citep{bluck14,bluck16,bluck20,teimoorinia16}.

AGN-driven jets are observed to influence the gaseous intracluster medium (ICM) associated with local galaxy clusters, inflating cavities that inhibit efficient radiative cooling \citep[e.g.][]{mcnamara07,fabian12,hlavacek15}, and help maintain the quiescence of the central galaxy. The effects of AGN feedback on the circumgalactic medium (CGM) of $\sim L^\star$ galaxies remain ill-constrained from both an observational and theoretical perspective \citep[for a review, see][]{tumlinson17}, but given the readily-observable effects of AGN feedback on the gas associated with more-massive haloes, it is plausible and arguably likely that AGN also significantly influence the properties of the CGM, and that the induced changes may play an important role in quenching.

In the absence of strong observational constraints on the connection between the properties of the CGM and the quenching of $\sim L^\star$ galaxies, one can instead seek insight from cosmological, hydrodynamical simulations in which the efficiencies of feedback processes are calibrated to ensure that the model produces realistic galaxies. The EAGLE simulations \citep{schaye15,crain15} represent an ideal testbed for such a study, since the parameters of their subgrid prescriptions for feedback are calibrated only against observations of well-characterised stellar properties of galaxies, leaving the nature of the gaseous universe as a prediction of the simulation. \citet{schaller15} showed that the EAGLE model expels a significant fraction of the baryons bound to dark matter haloes below the mass scales of galaxy groups, producing a monotonically-rising relation between the CGM mass fraction, $f_{\rm CGM}$, and halo mass. This expulsion is largely driven by the entrainment of CGM baryons in galactic outflows, since more gas is expelled from EAGLE haloes than is ejected from the interstellar medium of their central galaxies \citep{mitchell20}. 


Present-day galaxies in EAGLE exhibit significant diversity in $f_{\rm CGM}$ at a fixed halo mass, indicating differences in the impact of feedback on the gaseous environments of galaxies. \citet[][hereafter D19]{davies19} showed that at fixed halo mass, $f_{\rm CGM}$ correlates strongly with proxies for the halo assembly time, such that earlier-assembling haloes exhibit lower present-day CGM mass fractions than their later-assembling counterparts. D19 showed that this correlation is driven by AGN feedback; earlier-assembling haloes foster the growth of more massive central black holes (BHs) \citep{boothschaye10,boothschaye11}, which liberate more AGN feedback energy than is typical of BHs hosted by haloes of similar mass, thus expelling a greater fraction of the CGM. In a companion paper to D19, \citet{opp20a} showed that the most rapid phase of black hole growth is generally followed by a significant decrease in $f_{\rm CGM}$ for $\sim L^\star$ EAGLE galaxies, reflected clearly by decreasing covering fractions for ions such as H\textsc{i}, C\textsc{iv} and O\textsc{vi}. In tandem with baryon expulsion, AGN feedback also acts to reduce the baryon fraction of matter accreting onto $\sim L^\star$ EAGLE haloes \citep{wright20}, further suppressing $f_{\rm CGM}$.

Several studies undertaken with the EAGLE simulations have revealed correlations between assembly time, AGN feedback and properties of galaxies. Diversity in the star formation histories of galaxies, driven by differences in assembly time, appear to drive the scatter in the EAGLE stellar mass-halo mass relation \citep{matthee17} and star formation rate-stellar mass relation \citep{matthee19} in haloes of mass $M_{200}\lesssim 10^{12}\Msun$. At higher halo masses, galaxies with more massive central BHs (and hence earlier assembly times) exhibit lower star formation rates at fixed stellar and halo mass \citep[][D19]{matthee19}, and have redder colours \citep{opp20a}.

\citet[][hereafter D20]{davies20} demonstrated a mechanism through which these connections can arise: AGN-driven gas expulsion elevates the cooling time of the CGM, inhibiting the replenishment of the central galaxy's interstellar medium (ISM). This establishes a preference for early-assembling, gas-poor EAGLE haloes to host quenched, spheroidal/elliptical galaxies (and vice-versa). D20 showed that these correlations are also present in the IllustrisTNG simulations \citep[][hereafter TNG]{pillepich18b,nelson18a,springel18}, which employ markedly different subgrid treatments for both stellar and AGN feedback to those of EAGLE. Complementary results were also found by \citet{zinger20}, who examined the influence of AGN feedback on the thermodynamic state of circumgalactic gas at various radial distances from TNG galaxies.

These lines of evidence are suggestive of an intimate connection between the assembly histories of dark matter haloes, which are established by the initial power spectrum of density fluctuations (a property of the adopted cosmogony), and the astrophysical processes of quenching and morphological transformation. State-of-the-art galaxy formation models thus appear to indicate that the establishment of a red sequence of central galaxies requires that the content and structure of the CGM are transformed by AGN feedback in order to inhibit gas inflows, and that the impact of this effect (at fixed halo mass) is governed primarily by the assembly history of the parent dark matter halo. Galaxy formation models must self-consistently and realistically follow the evolution of both galaxies and their gaseous environments to capture this process, which appears to be a crucial and somewhat overlooked step in the quenching and morphological transformation of galaxies.

The large cosmic volumes followed by the current generation of state-of-the-art simulations of galaxy evolution yield populations of $\sim L^\star$ galaxies in diverse environments, enabling the comparison of similarly-massive haloes with markedly different assembly histories. A key limitation of this approach, however, is that these comparisons are necessarily made between {\it different} haloes, precluding the unambiguous establishment of a direct and exclusive causal connection between the properties of galaxy-CGM ecosystems and the assembly histories of their host dark matter haloes. Other potential driving factors, such as the environment of the halo for example, may also play a significant role. To remedy this, one might envisage performing simulations of increasingly large volume, enabling finer sub-sampling of the halo population to mitigate these effects, however this approach is clearly both costly and inefficient.

An alternative method, which yields a cleaner test of the influence of assembly history on galaxy properties, is to carefully modify the initial conditions of the matter comprising an individual halo, such that its assembly history can be adjusted whilst leaving the large-scale environment of the system unchanged, thus minimising other potential influences. Controlled experiments such as this can be realised through the use of the ``genetic modification" technique \citep[GM,][]{roth16,rey18}, an extension of the Hoffman-Ribak algorithm \citep{hoffman91}. Genetic modification yields an efficient and controlled means of examining the role of assembly history on the evolution of what is otherwise essentially the same central galaxy. 

We use the GM approach to examine, in a direct and systematic fashion, the influence of assembly history on the evolution of galaxy-CGM ecosystems. In a recent study,  \citet{sanchez19} used the GM technique to examine the influence of a dark matter merger history on the column density of circumgalactic \textsc{OVI}. Here, our focus is the role of assembly history in physically mediating galaxy evolution. We use the EAGLE galaxy formation model to simulate the assembly of a dark matter halo, whose present-day halo mass is (marginally) greater than the mass scale at which AGN feedback becomes efficient in the EAGLE model. We start from initial conditions that have been adjusted from the original `Organic' case to yield either accelerated or delayed halo assembly. We show that assembly history markedly influences the evolution of the central galaxy and its halo gas, and examine the response of the CGM to expulsive AGN feedback, showing that the reconfiguration of the CGM at lower density following AGN feedback events is key to facilitating quenching. Moreover, we show that the magnitude of these changes is significant when compared with the intrinsic uncertainty associated with the evolution of individual objects in galaxy formation models that use stochastic subgrid implementations of star formation and feedback. 

This paper is structured as follows: in Section \ref{sec:paper3:methods} we outline the rationale by which our candidate dark matter halo was selected and describe how the initial conditions were generated and genetically modified. We also give details of the simulation model, explain how the progenitors of the system are tracked through the simulation, and explain how certain diagnostic quantities were calculated. In Section \ref{sec:results:mod_histories} we evaluate the properties and assembly histories of the modified haloes, before examining the effects of a modified assembly history on the galaxy-CGM ecosystem in Section \ref{sec:results:eco}. Within the latter section, we examine the effects of assembly history on the properties of the central galaxy (Section \ref{sec:results:eco:gal}), on the growth of supermassive BHs and their associated AGN feedback (Section \ref{sec:results:eco:BH}), on the CGM mass fraction (Section \ref{sec:results:eco:CGM}), and on the structure and thermodynamic properties of the CGM (Section \ref{sec:thermo}). We summarise our results in Section \ref{sec:paper3:summary}. Throughout, we adopt the convention of prefixing units of length with `c' and `p' to denote comoving and proper scales respectively, e.g. ckpc for comoving kiloparsecs.

\section{Methods}
\label{sec:paper3:methods}

Our analyses are based on a suite of simulations that follow the formation and evolution of an individual $\sim L^\star$ central galaxy and its immediate environment, in its full cosmological context. This is most efficiently achieved via the adoption of `zoomed' initial conditions \citep[see e.g.][]{katz93,bertschinger01}, whereby only the object of interest is followed at high resolution and with hydrodynamics, whilst the remainder of the periodic volume is followed with purely collisionless dynamics and at reduced resolution. To evolve these initial conditions, we utilise the EAGLE version of the \textsc{gadget3} code. In this section we detail how the initial conditions of our simulations were generated and subsequently ``genetically modified" (Section \ref{sec:methods:ICs}). We then briefly describe the EAGLE model (Section \ref{sec:methods:eagle}), explain how we test for the influence of stochasticity on our results (Section \ref{sec:methods:seeds}), outline our techniques for identifying and characterising our galaxy-CGM system and its progenitors (Section \ref{sec:methods:halo}), and detail how various diagnostics used in our analysis were calculated (Section \ref{sec:methods:cooling}). We note that summaries of the EAGLE model are provided by many other studies, we therefore present only a brief description of the model in Section \ref{sec:methods:eagle}. Readers familiar with the model may wish to skip that section.

\subsection{Construction of the initial conditions}
\label{sec:methods:ICs}

To obtain a fiducial case of a present-day $\sim L^\star $ galaxy with a broadly typical specific star formation rate and circumgalactic gas fraction, we identify candidate \textit{galaxies} for resimulation from a parent volume evolved with a detailed galaxy formation model, rather than following the more common practice of identifying candidate dark matter haloes from a simulation evolved with purely collisionless dynamics. We identify candidate galaxies from a periodic simulation of uniform resolution whose initial conditions were generated with the \textsc{genetIC} software \citep{stopyra20}; use of this simulation rather than, for example, simulations from the EAGLE suite, simplifies the subsequent process of applying modifications to multi-resolution zoom initial conditions. This parent simulation adopts the best-fit cosmological parameters from \citet{planck16}, $h=0.6727$, $\Omega_{\rm 0}=0.3156$, $\Omega_{\Lambda} = 0.6844$, $\sigma_8=0.831$ and $n_{\rm s}=0.9645$. It is $L=50\cMpc$ on a side, and is populated with $N=512^3$ collisionless dark matter particles of mass $3.19 \times 10^7\Msun$ and an (initially) equal number of baryonic particles of mass $5.6 \times 10^6\Msun$. The cosmological parameters and particle masses are therefore sufficiently similar to those of the standard-resolution EAGLE simulations that simulating the volume with the EAGLE Reference model yields a galaxy population of similar realism to those simulations.

D20 noted that in both EAGLE and TNG, the influence of efficient AGN feedback on the mass fraction of the CGM, and by extension the properties of galaxies, is most apparent in haloes of present day mass $M_{200} \sim 10^{12.5}\Msun$. We therefore sought central galaxies hosted by haloes of this mass, and identified as our resimulation target a present-day central galaxy of stellar mass $M_{\rm \star}=4.3\times 10^{10}\Msun$, hosted by a halo of mass $M_{200}=3.4\times 10^{12}\Msun$ and virial radius $r_{200}=318\kpc$. The galaxy exhibits an extended stellar disc, has a stellar half-mass radius of $r_{\rm \star, 1/2} = 7.5$ kpc, and is steadily star-forming (sSFR = $10^{-10.2}\peryr$). The CGM mass fraction of the host halo, $f_{\rm CGM}$, normalised by the cosmic baryon fraction, $\Omega_{\rm b}/\Omega_{\rm 0}$, is 0.31. Following D20, we define $f_{\rm CGM}\equiv M_{\rm CGM}/M_{200}$, where $M_{\rm CGM}$ is the total mass of all gas within the virial radius that is not star-forming\footnote{This excludes the ISM, which is treated as a single-phase star forming fluid in the EAGLE model (See Section \ref{sec:methods:eagle})}. The centre of the nearest halo of at least equal mass is 3.3 Mpc from the centre of the target halo, a separation of more than 6 times the virial radius of the more massive halo; this ensures that the physical influence of neighbouring systems is negligible.
 
To construct the multi-resolution zoom initial conditions, we first identify all dark matter particles at $z=0$ residing within a sphere of radius $r=3r_{200}$, centred on the potential minimum of the halo, and trace them to their coordinates in the unperturbed (effectively $z=\infty$) particle distribution. The zoomed initial conditions of the halo are then constructed by masking the Lagrangian region defined by this particle selection in the unperturbed mass distribution; the enclosed mass distribution is resampled at higher resolution with a factor of 8 more particles (which represent both baryonic and dark matter), whilst the remainder of the volume is resampled with a factor of 8 fewer particles, which act as low-resolution boundary particles to provide the correct large-scale gravitational forces. The Zel'dovich displacements corresponding to the original phases and power spectrum (the latter now sampled to the higher and lower Nyquist frequencies of the high-resolution and boundary particles, respectively) are then reapplied. Evolution of these initial conditions yields the unmodified, or `Organic' assembly history.

We then apply the linear genetic modification technique of \citet{roth16} and \citet{pontzen17}\footnote{Further details of the technique are given by \citet{rey18}.} to construct a pair of complementary initial conditions, designed to yield modified halo assembly histories whilst maintaining approximately the same present-day halo mass. We have modified the initial conditions to yield assembly histories shifted, with respect to the Organic case, to both an earlier time (`GM-early') and a later time (`GM-late'). Specifically, the initial conditions were adjusted such that at $z=99$, the matter that will eventually comprise the halo's main progenitor at $z=2$ has a mean overdensity differing from the Organic case by factors of 1.05 and 0.95 respectively. The adjustments simultaneously ensure that the mean overdensity (at $z=99$) of the matter that will ultimately comprise the halo at $z=0$ remains unchanged, thus approximately fixing the final $z=0$ halo mass.

In common with the construction of the EAGLE initial conditions \citep[see Appendix B4 of][]{schaye15}, the final step (for both the Organic and modified initial conditions) is to replace the high-resolution particles in each case with a pair of particles consisting of a gas particle and a dark matter particle, with a gas-to-dark matter mass ratio of $\Omega_{\rm b} / (\Omega_{\rm 0} - \Omega_{\rm b})$. The masses of the gas, dark matter and low-resolution mass tracer particles are therefore $m_{\rm g} = 7.35 \times 10^{5}\Msun$, $m_{\rm dm} = 3.98 \times 10^{6}\Msun$, and $m_{\rm lr} = 3.02 \times 10^{8}\Msun$. The particle pairs are positioned such that their centre of mass corresponds to the position of the replaced particle, with the gas and dark matter particles moved in the ($1,1,1$) and ($-1,-1,-1$) coordinate directions, respectively.

\subsection{The EAGLE model}
\label{sec:methods:eagle}

We evolve the initial conditions to the present day with the EAGLE galaxy formation model \citep{schaye15,crain15,mcalpine16}. EAGLE uses the Tree-Particle-Mesh (TreePM) smoothed particle hydrodynamics (SPH) solver \textsc{gadget3} \citep[last described by][]{springel05}, with substantial modifications including the adoption of the pressure-entropy SPH formulation of \citet{hopkins13}, the artificial viscosity and artificial conduction switches of \citet{cullen10} and \citet{price10} respectively, and the time step limiter of \citet{durier12}. The subgrid physics implementation includes radiative heating and cooling for 11 individual elements \citep{wiersma09} in the presence of a time-varying UV/X-ray background \citep{haardtmadau01} and cosmic microwave background (CMB) radiation. The ISM is treated as a single-phase fluid with a polytropic pressure floor \citep{schaye08}. Gas with density greater than a metallicity-dependent threshold \citep{schaye04} is eligible for conversion to star particles stochastically \citep{schaye08}, at a rate that by construction reproduces the observed Kennicutt-Schmidt relation \citep{kennicutt98}. Star particles are treated as stellar populations with a \citet{chabrier03} stellar initial mass function (IMF), which evolve and lose mass according to the model of \citet{wiersma09b}, and inject feedback energy associated with star formation by stochastically and isotropically heating neighbouring gas particles by a temperature increment of $\Delta T_{\rm SF}=10^{7.5}\K$ \citep{dallavecchiaschaye12}. Black holes of initial mass $10^5$ M$_\odot/h$ are seeded on-the-fly at the centres of haloes with masses greater than $10^{10}\,M_\odot/h$ and act as ``sink'' particles that grow through BH-BH mergers and Eddington-limited Bondi-Hoyle accretion, modulated by the circulation speed of gas close to the BH \citep{rosasguevara15}. The associated feedback energy is injected by stochastically and isotropically heating neighbouring gas particles by a temperature increment of $\Delta T_{\rm AGN}=10^{8.5}\K$ \citep{boothschaye09,schaye15}. As motivated by \citet{schaye15} and described by \citet{crain15}, the efficiency of EAGLE's stellar feedback prescription was calibrated to reproduce both the present-day galaxy stellar mass function (GSMF) and the present-day sizes of galaxy discs, while the efficiency of the AGN feedback prescription was calibrated to reproduce the present-day relation between BH mass and stellar mass.

The particle mass resolution of our initial conditions ($M_{\rm DM}=3.98\times 10^6$ M$_\odot$) is slightly higher than the resolution at which the Reference EAGLE model was calibrated ($M_{\rm DM}=9.70\times 10^6$ M$_\odot$), resulting in a reduction of artificial radiative losses from gas heated by both stellar and AGN feedback. The galaxy population evolved with the EAGLE Reference model falls below the observed GSMF at the mass scale of our selected system \citep{schaye15}, indicating that feedback is too efficient on this scale; an increase in resolution will only exacerbate this issue \citep[see e.g.][]{font20}. We therefore adopt the recalibrated (RECAL) parameter values for EAGLE's subgrid feedback prescriptions, described by \citet[][their Table 3]{schaye15}, which yield an improved reproduction of the present-day GSMF at higher resolution. For each set of GM initial conditions, we have also run counterpart simulations in which no black holes are seeded and no AGN feedback occurs (NOAGN), and simulations considering only collisionless gravitational dynamics (DMONLY).

\subsection{Realisations with alternative random number seeds}
\label{sec:methods:seeds}
EAGLE's stochastic subgrid implementations of star formation and feedback require that a quasi-random number is drawn and compared to probabilities governed by local gas conditions. In the limit of adequate sampling, the influence of the intrinsic uncertainty associated with stochastic implementations diminishes, such that the properties of the galaxy population in a cosmic volume are, in a statistical sense, agnostic to the choice of the initial seed used by the quasi-random number generator. However, when considering the evolution of individual objects (as is the case here), this uncertainty can in principle be significant \citep{keller19}, and appears to be increasingly severe with decreasing resolution \citep{genel19}.

To assess the importance of this uncertainty for our zoom simulations, we evolve the three assembly histories nine times each, adopting each time a different seed for the quasi-random number generator used by the star formation and feedback routines. Comparison of the seed-to-seed scatter of a given property, for a fixed assembly history, enables us to assess whether differences in the same properties for simulations of different assembly histories are significant. Where we show a single reference case for a given assembly history, we choose the realisation that adopts the same seed value as the EAGLE simulations, but we note that this choice is not special, and that all nine realisations of each assembly history are equivalent.

\subsection{Identifying haloes, galaxies and their progenitors}
\label{sec:methods:halo}

Haloes are identified in the simulations through the application of the friends-of-friends (FoF) algorithm to the dark matter distribution, with a linking length of 0.2 times the mean interparticle separation. Gas, star and BH particles are then assigned to the FoF group (if any) of their nearest dark matter particle. The SUBFIND algorithm \citep{springel01,dolag09} is then used to identify bound substructures within haloes. Throughout this work, the properties of haloes, such as the spherical overdensity mass ($M_{200}$) and CGM mass fraction ($f_{\rm CGM}$), are computed within a radial aperture $r_{200}$, centred on the halo's most bound particle, which encloses a mean density equal to 200 times the critical density, $\rho_{\rm crit}$. Galaxy properties, such as the specific star formation rate, are computed by aggregating the properties of the relevant particles within 30pkpc of the halo centre, following \citet{schaye15}. 

To obtain the merger history of the genetically-modified galaxy-halo system, we first identify the 100 most bound dark matter particles\footnote{The recovered merger history is not strongly sensitive to this choice.} comprising the halo in the Organic case, realised using the standard EAGLE random number seed, at $z=0$. The main progenitor of the halo in each prior snapshot is then defined as the subhalo containing the greatest fraction of these particles, thus yielding a reference merger history. To identify the equivalent structure at the same epoch in other simulations, i.e. those with modified assembly histories, or those employing different physical models (NOAGN/DMONLY), we cross-match the 100 most-bound dark matter particles comprising the main progenitor at that epoch. This method affords a reliable means of tracking of the same object across all simulations. In the very early stages of the halo's assembly, poor particle sampling of the halo, combined with differences in accretion history between the Organic and GM systems, can result in the misidentification of the correct progenitor in the GM-early and GM-late simulations. We highlight where this is the case throughout via the use of dotted curves.

\begin{figure*}
\includegraphics[width=\textwidth]{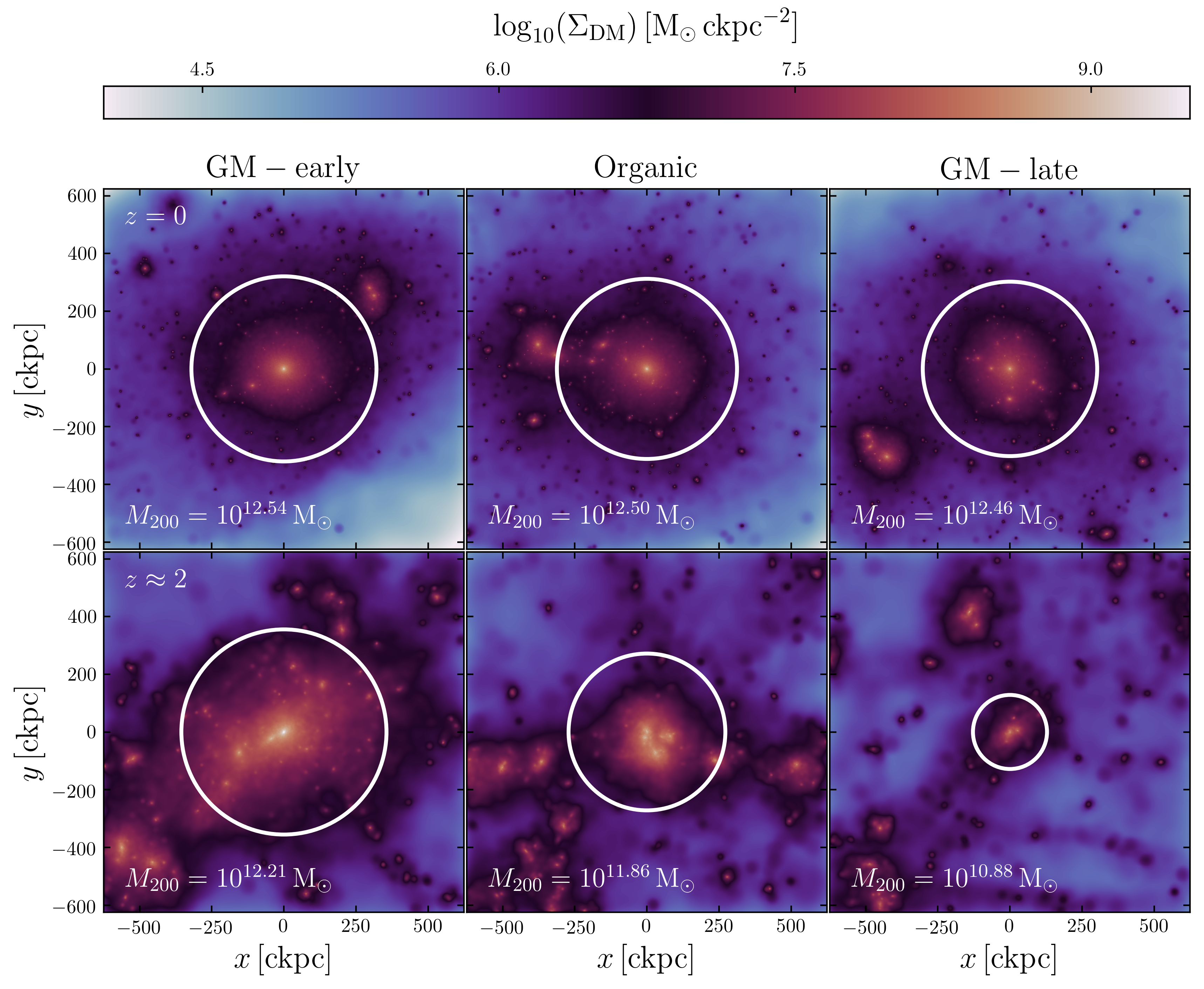}
\caption{Surface density maps of the dark matter distribution for the GM-early (left column), Organic (centre column) and GM-late (right column) haloes at $z=0$ (upper row) and at $z\approx 2$ (lower row). The white circle on each panel denotes the virial radius, $r_{200}$. The field of view in each case is $1.28\cMpc$, which corresponds to $4r_{200}$ for the Organic assembly history at $z=0$. The halo mass, $M_{200}$, is quoted on each panel. The haloes exhibit similar present-day $M_{200}$, but have significantly different masses at $z\approx 2$ as a result of their differing assembly histories.}
\label{fig:DMimage}
\end{figure*}

\subsection{Feedback energetics and cooling timescales}
\label{sec:methods:cooling}

We use the total energy injected by AGN feedback relative to the binding energy of the halo baryons as a diagnostic quantity in Section \ref{sec:results:eco:BH}. The total energy injected over the lifetime of a BH of mass $M_{\rm BH}$ at time $t$ is given by:
\begin{equation}
    E_{\rm AGN}(t) = \frac{\epsilon_{\rm f}\epsilon_{\rm r}}{1-\epsilon_r}(M_{\rm BH}(t)-M_{\rm BH,seed}) c^2,
\label{eq:e_agn}
\end{equation}
where $M_{\rm BH,seed}=10^5$ M$_{\odot}/h$ is the seed mass of the BH, $c$ is the speed of light, and $\epsilon_{\rm r}=0.1$ is the radiative efficiency assumed for the BH accretion disk. The parameter $\epsilon_{\rm f}=0.15$ specifies the fraction of the radiated energy that couples to the surrounding gas, and its value is calibrated to reproduce the observed relation between the BH mass and the galaxy stellar mass \citep{schaye15}. Approximately $1.67\%$ of the rest mass energy of gas accreted by the BH is therefore coupled to the gas surrounding the BH. We subtract the contribution from the BH's seed mass, since it has not been injected into the gas in the simulation. We note that this definition includes the energy injected by progenitor BHs that have merged into the descendant BH; we do not subtract this contribution since it directly affects the CGM of the descendant galaxy.

The ``intrinsic'' binding energy of the baryons at time $t$, $E_{\rm bind}^{\rm b}(t)$, is obtained by calculating the binding energy of the halo in an equivalent, collisionless, dark-matter only simulation\footnote{As discussed by D19, we use intrinsic binding energy measurements from the DMONLY simulation, because the inclusion of baryonic physics can systematically alter the binding energy of the underlying dark matter structure to a degree comparable with the intrinsic scatter at a given $M_{200}$.}, and multiplying this by the cosmic baryon fraction: $E_{\rm bind}^{\rm b}(t)=(\Omega_{\rm b}/\Omega_0)E_{\rm DMO}^{200}(t)$. 

We examine the radiative cooling times of circumgalactic gas particles in Section \ref{sec:thermo}, which we compute for sets of particles by dividing the sum of their total internal thermal energies, $u_i$, to their total bolometric luminosities, $L_{{\rm bol},i}$, via $t_{\rm cool} = \sum_i{u_i}/\sum_i{L_{{\rm bol},i}}$. The bolometric luminosity is given by $L_{{\rm bol},i} = n_{{\rm H},i}^2\Lambda_i V_i$, where $n_{{\rm H},i}$ is the hydrogen number density of the gas particle and $V_i=m_{{\rm g},i}/\rho_i$ where $m_{{\rm g},i}$ is the particle mass and $\rho_i$ is its mass density. $\Lambda_i$ is the particle's volumetric net radiative cooling rate (i.e. incorporating radiative cooling plus photoheating) specified by its temperature, density, element abundances, and the incident flux from the CMB and metagalactic UV/X-ray radiation fields.

Consistent with the implementation of radiative cooling in EAGLE, we use the volumetric net radiative cooling rates tabulated by \citet{wiersma09}, which were computed using \textsc{CLOUDY} version 07.02 \citep{ferland98}. The tables specify the cooling rate as a function of hydrogen number density, $n_{\rm H}$, temperature, $T$, and redshift, $z$ for each of the 11 elements tracked by the EAGLE model (H, He, C, N, O, Ne, Mg, Si, S, Ca and Fe), and we interpolate them in $\log_{10} n_{\rm H}$, $\log_{10} T$, $z$, and, in the case of the metal-free cooling contribution, the helium fraction $n_{\rm He}/n_{\rm H}$. The element-by-element contributions are then used to compute the net cooling rate for the particle:
\begin{equation}
\label{eq:chap5:coolrate}
    \Lambda = \Lambda_{\rm H,He} + \sum_{i>{\rm He}} \Lambda_{i,\odot} \frac{n_e/n_{\rm H}}{(n_e/n_{\rm H})_{\odot}} \frac{n_i/n_{\rm H}}{(n_i/n_{\rm H})_{\odot}},
\end{equation}
where $\Lambda_{\rm H,He}$ is the metal-free contribution, $\Lambda_{i,\odot}$ is the contribution of element $i$ for the solar abundances assumed in \textsc{CLOUDY}, $n_e/n_{\rm H}$ is the particle electron abundance, and $n_i/n_{\rm H}$ is the particle abundance in element $i$. 

\section{Genetically modified assembly histories}
\label{sec:results:mod_histories}

We begin with an examination of the assembly histories and present-day properties of the haloes yielded by our application of the genetic modification technique. Recall that our aim is to systematically shift the assembly of the halo to earlier or later times, without inducing strong changes to the mass of the halo at $z=0$. Fig.~\ref{fig:DMimage} shows maps of the dark matter surface density of the GM-early (left column), Organic (centre column) and GM-late (right column) haloes in the RECAL simulations, at the present day (upper row) and at $z=2$ (lower row). The white circle on each panel denotes the virial radius, $r_{200}(t)$. The field of view in each case is $1.28\cMpc$, which corresponds to $4r_{200}$ for the Organic halo at $z=0$. The images present a striking representation of the effect of the technique: the structure of the Organic and modified haloes at the present day is similar, but at $z=2$ major differences in the structure of the halo's main progenitor are evident, with the assembly of the GM-early (GM-late) case being significantly advanced (delayed) with respect to the Organic case. 

\begin{figure}
\includegraphics[width=\columnwidth]{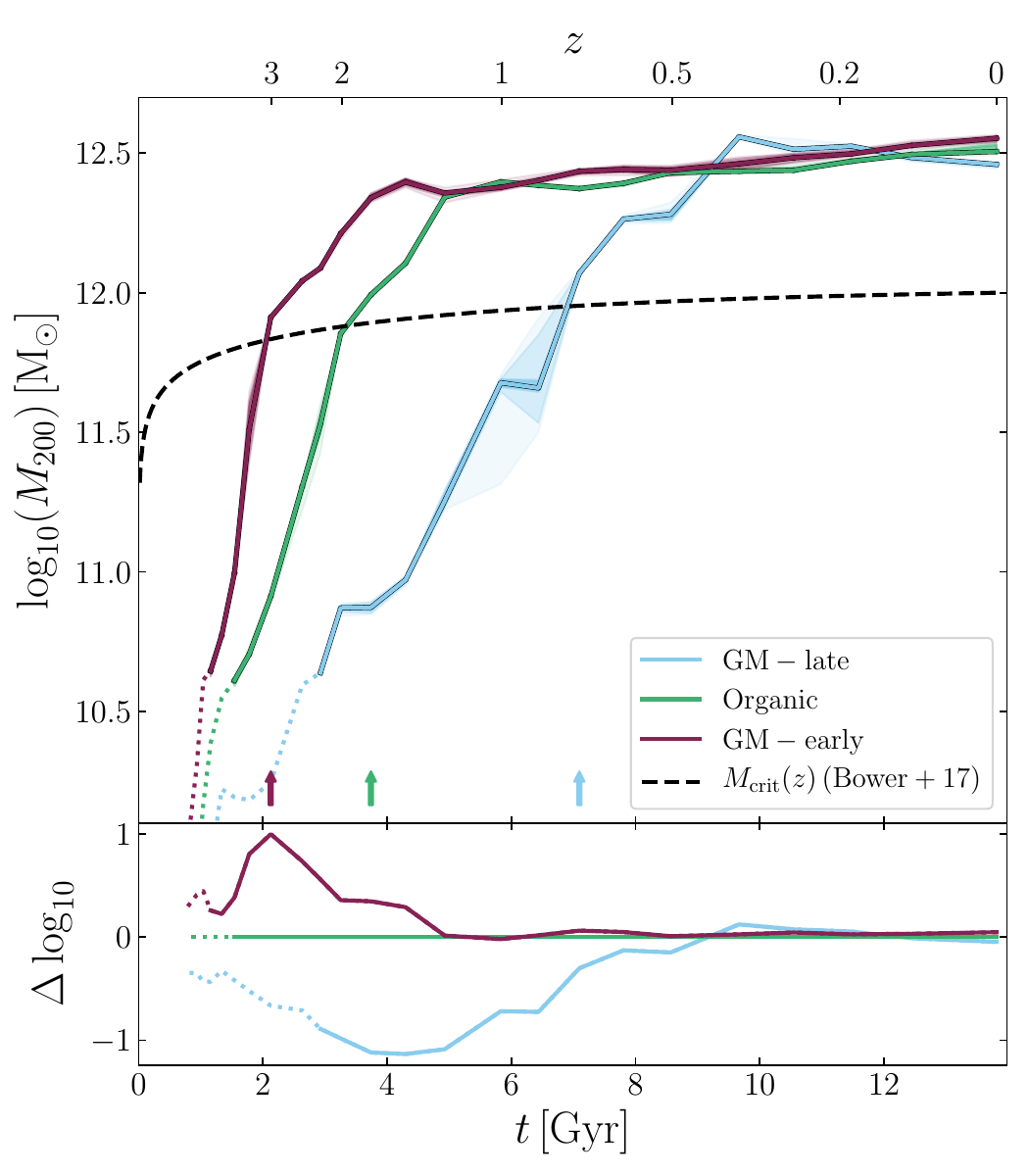}
\caption{The halo mass accretion histories, $M_{200}(t)$ of the three families of RECAL simulations. The three families attain similar final halo masses, but have markedly different accretion histories. Solid lines show the median $M_{200}$ of the nine simulations of each accretion history, with each run adopting a different seed for the quasi-random number generator used by EAGLE's stochastic implementations of star formation and feedback. The distribution across the nine simulations of each history is denoted by progressively lighter shading between pairs of seeds that give increasingly divergent results from the median. Evolutionary tracks are shown as dotted lines where $M_{200}(t)/M_{200}(z=0)<0.01$. The sub-panel shows the deviation, $\Delta(t)$, of ($\log_{10}$ of) the median halo mass of each family with respect to that of the Organic family. The black dashed line denotes the redshift-dependent critical halo mass, $M_{\rm crit}(z)$, where the buoyant transport of outflows driven by stellar feedback in EAGLE is expected to cease being efficient \citep{bower17}. The time corresponding to the first snapshot output for which $M_{200}(t)>M_{\rm crit}(z)$ is denoted by a coloured arrow for each family of simulations.}
\label{fig:ev:m200}
\end{figure}

Fig.~\ref{fig:ev:m200} shows the mass accretion history of the haloes, i.e. the halo mass of the main progenitor of the present-day halo, $M_{200}(t)$. As noted in Section \ref{sec:methods:halo}, the identification of the main progenitor can be ambiguous at early times, therefore the tracks in each case are plotted with dotted lines until $M_{200}(t)/M_{200}(z=0) = 0.01$. Here, and in subsequent figures of this style, for each of the GM-early, Organic and GM-late families we show the evolution derived from all nine of the simulations run in each case (each adopting a different initial seed for the quasi-random number generator). The solid lines do not represent the evolution of the system for any particular seed, but rather show the median value of the quantity of interest at each epoch. The differences in the median from that of the Organic family is shown in the sub-panels of all plots of this type. The distribution of values emerging from the nine simulations comprising each of the GM-early, Organic and GM-late families is illustrated with progressively lighter shading between pairs of seeds that give increasingly divergent results from the median value. We quantify this scatter with the interquartile range (IQR); in a rank-ordered sample of values taken from the nine simulations comprising each family, the third and seventh values are good approximations for the 25$^{\rm th}$ and 75$^{\rm th}$ percentiles, and we quote the difference between these as the IQR throughout.

We also show as a dotted line on Fig.~\ref{fig:ev:m200} the redshift-dependent critical halo mass introduced by \citet{bower17}, namely the mass at which buoyant transport of winds from EAGLE's stellar feedback is expected to cease to be efficient, $M_{\rm crit}=10^{12} (\Omega_0(1+z)^3 + \Omega_{\Lambda})^{-1/8}\,M_\odot$. \citet{bower17} noted that the rapid growth of central BHs in EAGLE tends to begin when haloes reach this mass, and interpreted this as a signature of BH fuelling by cooling flows from the quasi-hydrostatic CGM that builds in response to the cessation of buoyant transport. The epoch at which this threshold is reached by the median of each of family is denoted by an arrow and has value $t=2.13$ Gyr (GM-early), $t=3.74$ Gyr (Organic) and $t=7.10$ Gyr (GM-late). We discuss the consequences of the significant difference of these values in following sections.

As is critical for our purposes, the genetic modification process induces strong deviations from the Organic assembly history in the modified cases. At $z=2$, the GM-early system has already reached a halo mass of $\log_{10}(M_{200}/M_\odot)=12.21$ (IQR$=0.01$ dex), while the Organic halo has a mass of $\log_{10}(M_{200}/M_\odot)=11.86$ (IQR$=0.01$ dex) and the GM-late halo has assembled a halo mass of only $\log_{10}(M_{200}/M_\odot)=10.87$ (IQR$=0.01$ dex). These values represent 47 percent, 23 percent and 3 percent of the final halo masses, respectively. By $z=1$, the evolutionary tracks of the GM-early and Organic cases converge, and the halo mass evolves in a similar fashion for both thereafter. In the GM-late simulations, the halo mass evolves much more steadily, only attaining (and briefly exceeding) the mass of the other realisations after $z=0.5$. Despite these significant differences in mass accretion history, the present-day halo masses of the haloes are very similar, at $\log_{10}(M_{200}/M_\odot)=12.54$ (GM-early, IQR$=0.01$ dex), 12.50 (Organic, IQR$=0.02$ dex) and 12.46 (GM-late, IQR$=0.01$ dex). As might be expected, there is little scatter between runs adopting different seeds.

We contextualise these assembly histories by comparing them to those of haloes with similar present-day masses in the largest EAGLE simulation volume (Ref-L100N1504). We construct a sample of such haloes in a 0.2 dex wide window about $M_{200}=10^{12.5}\Msun$ and trace them back to their main progenitors at $z=2$. At this epoch, the median halo mass of the GM-early family of simulations is shifted from the median $M_{200}$ of this sample by $+1.9\sigma$, while the median GM-late halo mass is shifted by $-1.7\sigma$. The mass of the Organic halo lies close to the median $M_{200}$ of the sample at $z=2$, deviating by only $-0.2\sigma$. The modified histories therefore differ significantly from the typical assembly history of a halo of this mass, but are not extreme cases.

Differences in the collapse time of dark matter haloes (of a fixed present-day mass) yield differences in their density profiles, and hence in their concentrations and binding energies \citep[e.g.][]{neto07}; one should therefore expect that our modified haloes have differing concentrations. To examine the intrinsic differences arising solely from assembly history (i.e. in the absence of baryonic processes), we compute the $z=0$ concentration, $c_{\rm DMO}$, of each halo's counterpart in the DMONLY simulations by fitting an NFW profile following the method of \citet{neto07}. The GM-early system exhibits the highest concentration, $c_{\rm DMO}=9.1$, followed by the Organic ($c_{\rm DMO}=6.6$) and GM-late ($c_{\rm DMO}=5.7$) systems.

The above results demonstrate that the genetic modification technique enables controlled, systematic adjustment of the assembly history of an individual system, while only introducing changes of $\pm 0.04$ dex in the final halo mass. Note that these differences are dependent on the precise definition of halo mass (e.g. $M_{200}$, $M_{500}$ etc), since the altered accretion history also necessarily changes the density profile of the halo. While the GM technique permits iterative adjustment to match any particular definition of the final halo mass more precisely \citep{rey18}, this would not serve any particular physical purpose in the context of our numerical experiments. In the Ref-L100N1504 EAGLE simulation, a shift of 0.1 dex about $M_{200}=10^{12.5}$ M$_\odot$ corresponds to a difference in the median $f_{\rm CGM}$ of only 0.028. Greater differences in $f_{\rm CGM}$ can therefore be reasonably interpreted as a response to the modified assembly history of the halo. 

\section{The influence of halo assembly history on the galaxy-CGM ecosystem}
\label{sec:results:eco}

We turn now to the properties of the galaxies realised with each set of initial conditions in Section~\ref{sec:results:eco:gal}. To investigate the differences induced by adjustment of the halo assembly history, we then examine the evolution of the central SMBH mass in Section~\ref{sec:results:eco:BH}, the evolution of the CGM mass fraction in Section~\ref{sec:results:eco:CGM} and the structure and properties of the CGM in Section \ref{sec:thermo}.

\subsection{Properties of the central galaxy}
\label{sec:results:eco:gal}

\begin{figure}
\includegraphics[width=\columnwidth]{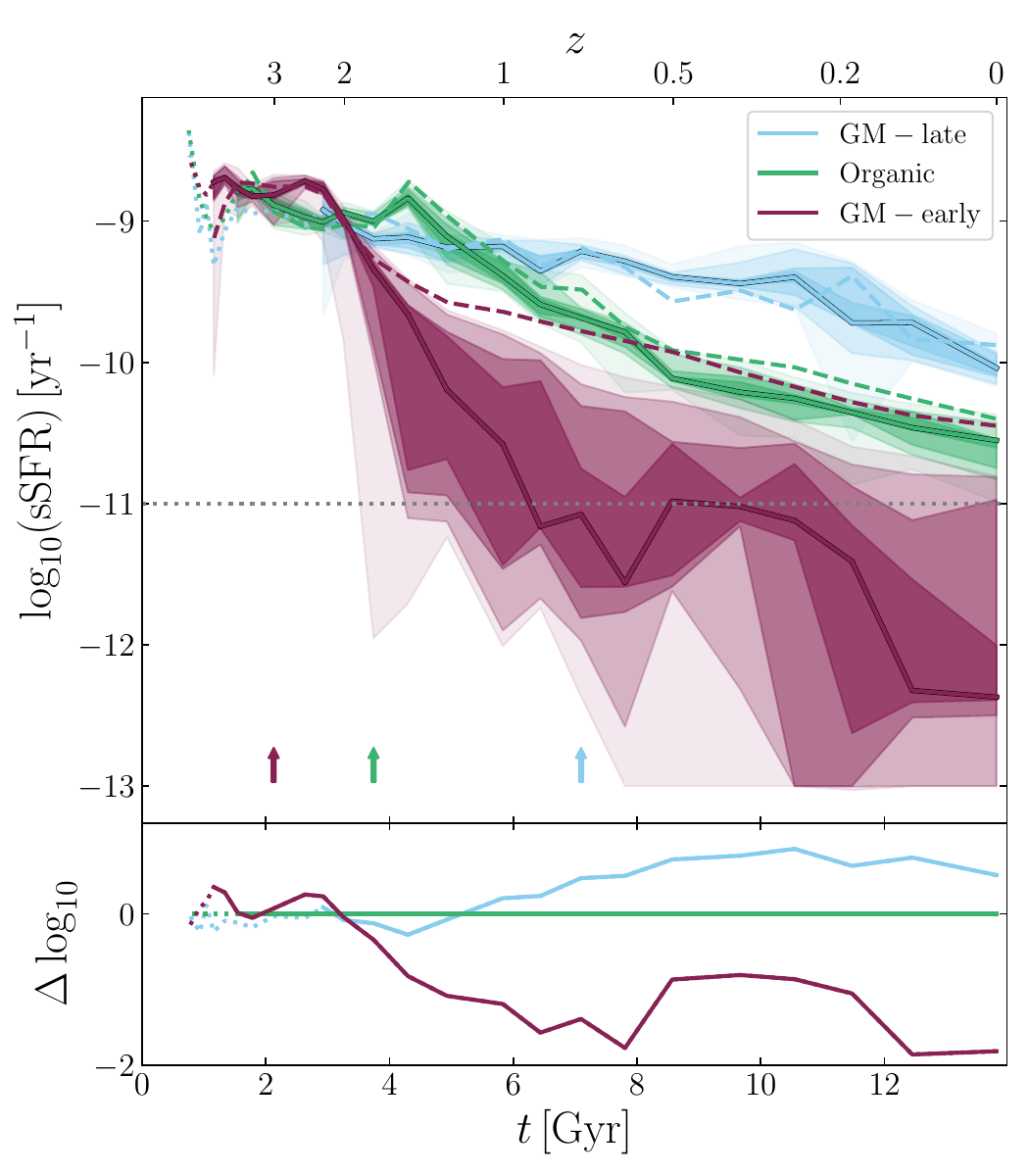}
\caption{The evolution of the specific star formation rate (sSFR) of the central galaxy for the three families of simulations. Curves, shading and arrows are defined as per Fig. \ref{fig:ev:m200}. Dashed lines show the median sSFR in simulations where AGN feedback is disabled (i.e. where no BHs are seeded). A minimum value of $10^{-13}$ yr$^{-1}$ is imposed for clarity, and we denote the canonical threshold for quenching, sSFR$=10^{-11}$ yr$^{-1}$, with a grey dotted line. The sub-panel show the deviation, $\Delta(t)$, of ($\log_{10}$ of) the median sSFR of each family from that of the Organic family. Earlier halo assembly leads to a stronger reduction in the sSFR; all simulations in the Organic and GM-late families yield present-day star forming galaxies, while all but three of the GM-early simulations yield passive galaxies at $z=0$.}
\label{fig:ev:ssfr}
\end{figure}

\begin{figure}
\includegraphics[width=\columnwidth]{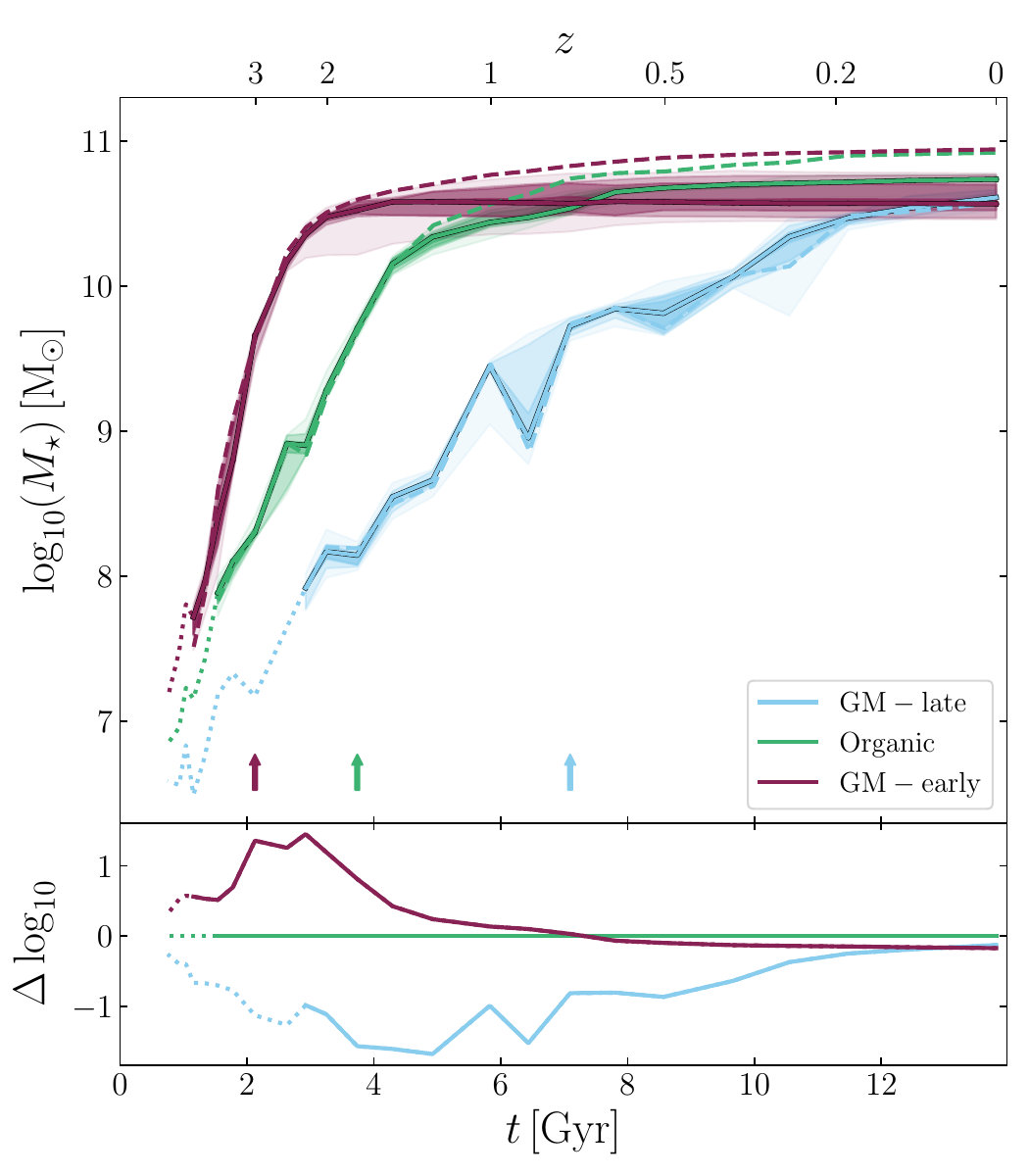}
\caption{The evolution of the central galaxy stellar mass, $M_\star(t)$, for the three families of simulations, shown in the same fashion as Fig. \ref{fig:ev:ssfr}. Dashed lines show the median stellar mass in simulations where AGN feedback is disabled. Sub-panels show the deviation, $\Delta(t)$, of ($\log_{10}$ of) the median stellar mass from that of the Organic case. The three families attain similar final stellar masses, but have markedly different stellar mass assembly histories.}
\label{fig:ev:mstar}
\end{figure}

We first explore how differences in the halo's mass accretion history influence the evolution of the central galaxy. Fig.~\ref{fig:ev:ssfr} shows the evolution of the sSFR of the central galaxy of the halo in the three families. The sSFR is computed at each snapshot epoch, averaged over the preceding 300 Myr to suppress sampling noise in, and short-term variation of, the star formation rate \citep[which can vary significantly on short, $\sim 10^6$ yr timescales, see][]{mcalpine17}. For clarity, we impose a minimum value of $10^{-13}\peryr$, and note that in any case it is not feasible to infer lower, non-zero values from observational measurements. We note that the evolution of the sSFRs of the EAGLE galaxy population was explored in detail by \citet{furlong15}.

The sSFR decreases with advancing time in all three families of simulations. In the Organic and GM-late cases, the decline is relatively shallow, and in all simulations from these families the galaxy remains actively star forming at $z=0$, with medians of ${\rm sSFR}>10^{-10.6}\peryr$ and ${\rm sSFR}=10^{-10.0}\peryr$, respectively. In contrast, the median sSFR for the GM-early family declines rapidly at $z\approx 2$, effectively `quenches' (i.e. its sSFR drops below $10^{-11}\peryr$) at $z=0.86$, and remains quenched at all subsequent times, with median ${\rm sSFR}=10^{-12.4}\peryr$ at $z=0$. The scatter between runs of differing random seed values is mild for the Organic and GM-late cases, with present-day IQR values of 0.2 dex, but is much more significant for the GM-early case, at $1.5\dex$, driven primarily by the sampling noise that arises at low SFRs due to the discreteness of the forming star particles. In consequence, three of the nine simulations from this family yield galaxies that remain star forming (i.e. sSFR$> 10^{-11}\peryr$) at $z=0$. This scatter notwithstanding, it is clear that the median trends of the three families are unambiguously influenced by the halo assembly history. By adjusting this property for an individual halo, we are able to convert a star-forming galaxy with a typical present-day star formation rate into either a more vigorously star-forming galaxy, or one that is quenched.

To demonstrate that AGN feedback is essential to the quenching of the galaxy, we add dashed lines to Fig.~\ref{fig:ev:ssfr} to show the median sSFR evolution for each family in simulations with AGN feedback disabled (i.e. where no BHs are seeded). There is little difference for the GM-late case, indicating that AGN feedback does not significantly affect the evolution of the galaxy for this assembly history, and in the Organic case disabling the AGN feedback only mildly elevates the sSFR for $z \lesssim 1.5$. Conversely, disabling AGN feedback in the GM-early case results in the galaxy never quenching, retaining a present-day sSFR of $10^{-10.4}\peryr$.

We now examine the impact of these differences in star formation history by showing the evolution of the stellar masses of the central galaxy for the three families of simulations in Fig. \ref{fig:ev:mstar}. The galaxies attain similar present-day stellar masses: $\log_{10}(M_\star/M_\odot)=10.57$ (GM-early, IQR$=0.20$ dex), 10.74 (Organic, IQR$=0.01$ dex) and 10.61 (GM-late, IQR$=0.07$ dex), but exhibit markedly different stellar mass assembly histories, with that of GM-early (GM-late) being significantly advanced (delayed) with respect to the Organic case. In all three cases, the stellar mass assembly initially occurs in concert with the halo mass assembly (cf. Fig. \ref{fig:ev:m200}), indicating that differences in the `cosmological' accretion rate sets the availability of gas for star formation in the early stages of the galaxy's formation. In the GM-early case, however, the median stellar mass of the galaxy reaches a maximum value at $z\approx 1.3$, shortly after its sSFR begins to rapidly decline (cf. Fig. \ref{fig:ev:ssfr}). The median stellar mass of the GM-early galaxy steadily declines due to stellar mass loss thereafter, while the Organic and GM-late galaxies remain star-forming and overtake the GM-early $M_\star$ at $z\approx 0.6$ and $z=0$ respectively. 

As for Fig. \ref{fig:ev:ssfr}, we also show the median $M_\star$ evolution for each family in simulations with AGN feedback disabled using dashed lines. In the absence of AGN feedback, the stellar mass growth of the GM-early galaxy is not halted and continues to the present day, yielding a present-day median stellar mass of $\log_{10}(M_\star/M_\odot)=10.94$. The present-day Organic stellar mass is higher in the absence of AGN, with a median of $\log_{10}(M_\star/M_\odot)=10.92$, while the stellar mass growth of the GM-late galaxy is relatively unaffected by AGN feedback. 

Figures \ref{fig:ev:m200}-\ref{fig:ev:mstar} demonstrate that AGN feedback is primarily responsible for the significant differences in present-day sSFR exhibited by our three systems. From $z\approx 1.3$ until the present day, the median halo mass accretion rates of the GM-early and Organic systems are very similar, and under the assumption that the accreted material has a constant baryon fraction, the gas infall rates should therefore also be similar. Throughout this period, however, the central galaxy of the Organic system is actively star-forming, while the GM-early galaxy is quenched for the majority of this time and its stellar mass does not increase. In the absence of AGN feedback, the GM-early galaxy remains star-forming and the difference between the median sSFR of this family and that of the Organic family is greatly reduced, therefore this source of feedback must be primarily responsible for preventing the gas inflowing onto the GM-early halo from fuelling star formation in its central galaxy \citep[see also][]{vandevoort11a,vandevoort11b}.

\begin{figure}
\includegraphics[width=\columnwidth]{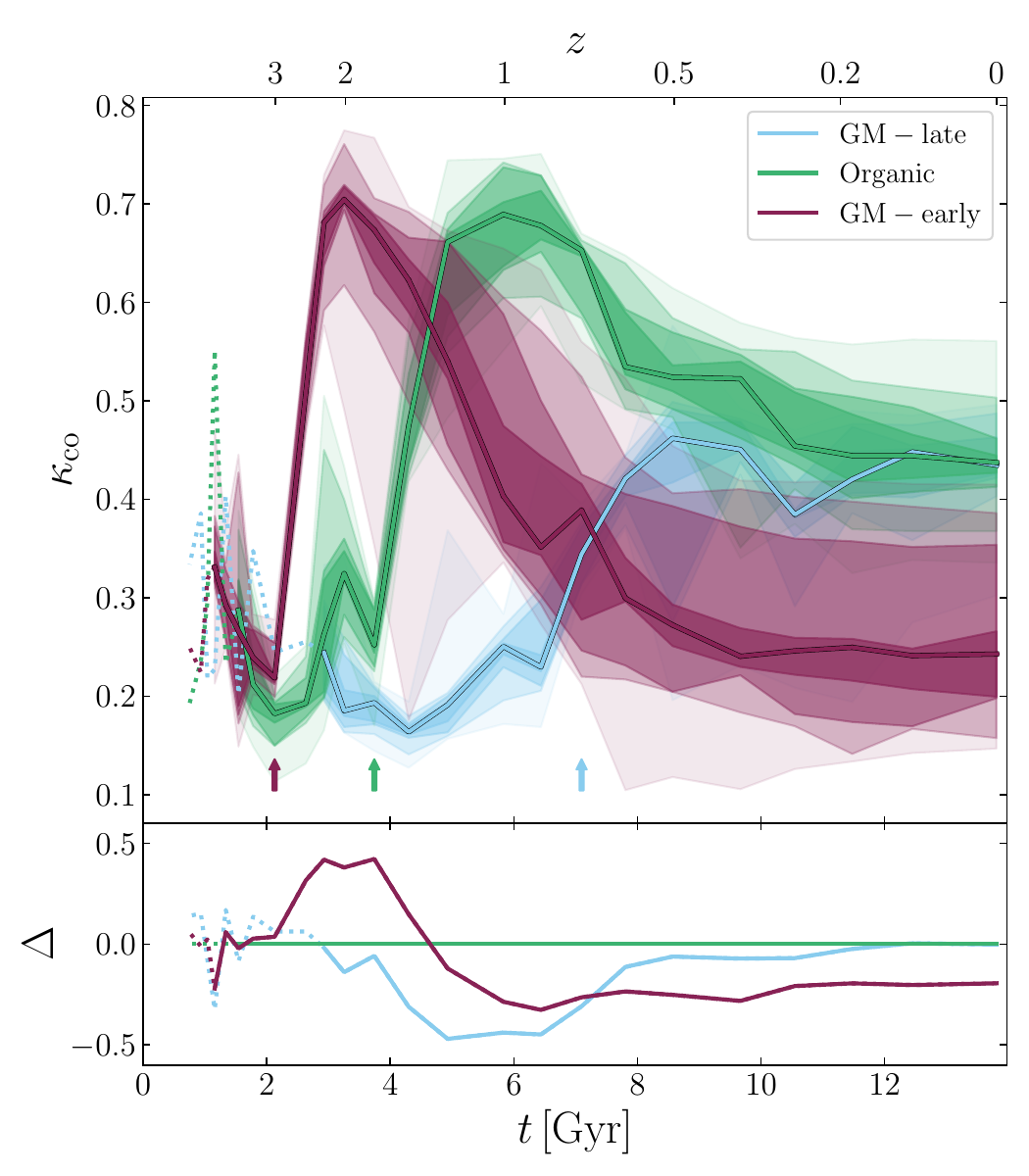}
\caption[The evolution of the central galaxy $\kappa_{\rm co}$ for the three families of simulations.]{The evolution of the stellar co-rotational kinetic energy fraction, $\kappa_{\rm co}$, of the central galaxy for the three families of simulations, shown in the same fashion as Fig. \ref{fig:ev:m200}. The sub-panel shows the deviation, $\Delta(t)$, of the median value of each family from that of the Organic family. The Organic and GM-late assembly histories yield present-day disc galaxies with $\kappa_{\rm co}>0.4$, while the simulations of the GM-early assembly history yield galaxies with spheroidal morphology, i.e. $\kappa_{\rm co}<0.4$.}
\label{fig:ev:kappa}
\end{figure}

Quenched galaxies that populate the red sequence are typically observed to have spheroidal morphologies and predominantly dispersive kinematics \citep[e.g.][]{kelvin14,vandesande17,vandesande18}. We therefore now examine how adjustments to the halo assembly history impact the kinematical evolution of the central galaxy. Fig.~\ref{fig:ev:kappa} shows the evolution of the stellar co-rotational kinetic energy fraction, \kappaco\footnote{We compute \kappaco using the publicly-available routines of \citet{thob19}.}, of the main progenitor of the central galaxy, for the three families of simulations. \kappaco is defined as the fraction of the kinetic energy of the stars in the galaxy that is invested in co-rotational motion, for which a threshold value of 0.4 has been shown to separate star-forming discs ($\kappa_{\rm co}>0.4$) from quenched spheroids ($\kappa_{\rm co}<0.4$) in EAGLE \citep{correa17}. 

For all three assembly histories, the kinematics of the central galaxy initially exhibits little co-rotational motion, before \kappaco rises to maximal values of $\kappa_{\rm co}=0.70$ (GM-early), $\kappa_{\rm co}=0.69$ (Organic) and $\kappa_{\rm co}=0.46$ (GM-late) at $z=2$, $z=1$ and $z=0.5$ respectively, signalling the formation of a rotationally-supported disc. Comparison of the evolution in \kappaco with that of $M_{200}$ in Fig. \ref{fig:ev:m200} illustrates that the difference in timing is due to the halo mass assembly history; the majority of the co-rotational kinetic energy in the galaxy is built up over the period in which the host halo accretes most of its final mass. For all seed values, the GM-late galaxy then retains a similar value of \kappaco until the present day, while \kappaco declines for the Organic and GM-early galaxies after reaching a maximum. At the present day, the GM-late and Organic galaxies are disc-like (i.e. $\kappa_{\rm co}>0.4$), with similar median values of $\kappa_{\rm co}=0.43$ and $\kappa_{\rm co}=0.44$ respectively. By this same definition, the GM-early galaxy becomes spheroidal at $z\simeq 1$, with a median across the nine seed values of $\kappa_{\rm co}=0.24$ at $z=0$.

The scatter induced by using different random number seed values is mild for the GM-late and Organic cases, with present-day IQR values of 0.05 and 0.04 respectively, but it is stronger for the GM-early case (IQR$=0.16$). D20 reported that the correlations between $\kappa_{\rm co}$ and properties of the halo (such as $f_{\rm CGM}$) are weaker than for the sSFR, and the overlap in the scatter for the three cases here reflects this. Nonetheless, for the GM-early assembly history, eight of the nine seeds yield a galaxy with $\kappa_{\rm co}<0.4$ at the present day. The morphologies of galaxies can clearly be changed from disc-like to spheroidal through a controlled modification of the halo assembly history.

These results are complementary to those of D20, who showed that in EAGLE and TNG, there exists a significant correlation (at fixed halo mass) between proxies for halo assembly time and the degree of rotational support in the stellar kinematics of central galaxies. Here we demonstrate a causal connection between these quantities, since our controlled experiment enables us to compare directly the morphological evolution of individual galaxies that differ only in the assembly histories of their host haloes.

\begin{figure*}
\includegraphics[width=\textwidth]{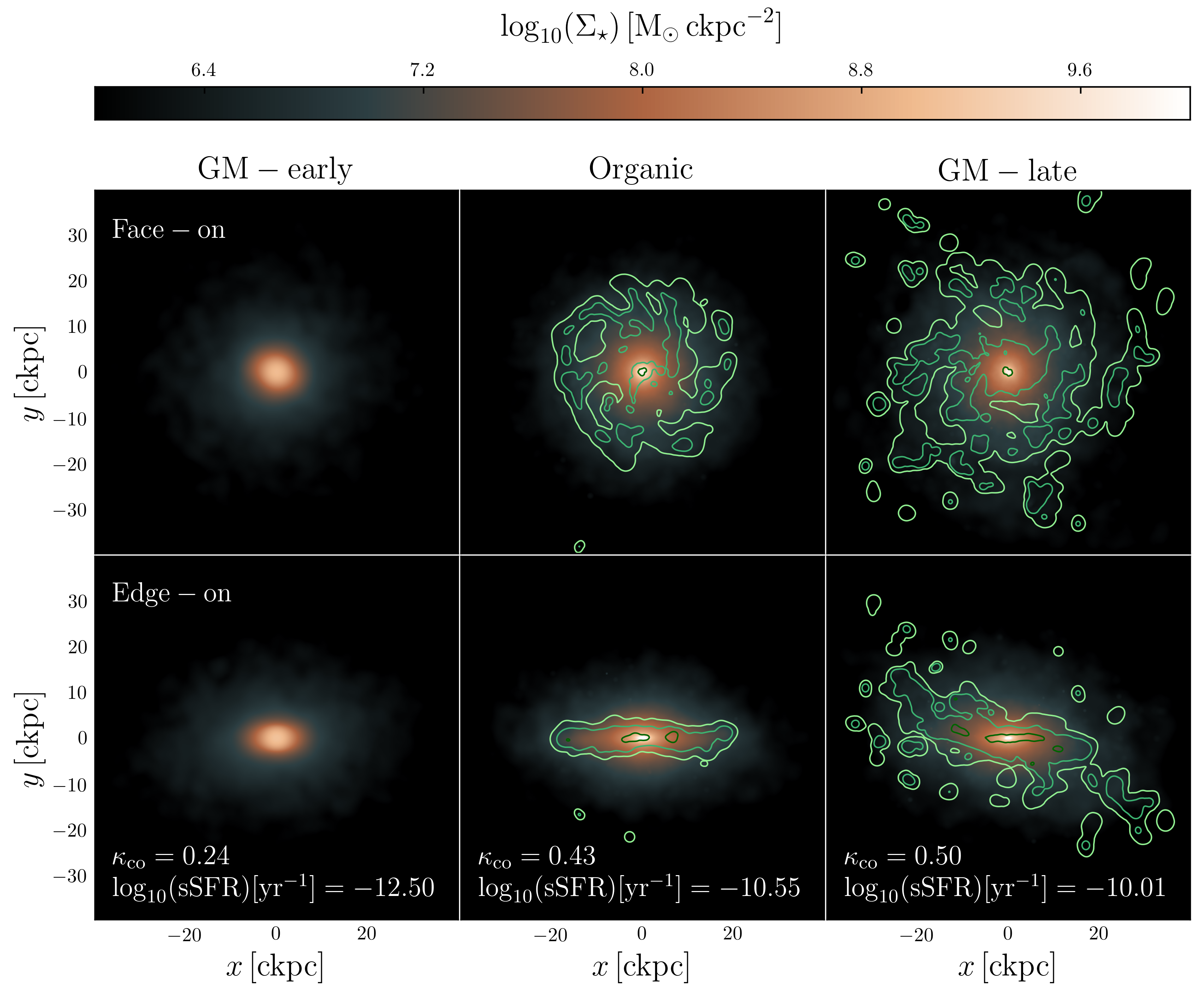}
\caption{Face-on (upper row) and edge-on (lower row) present-day surface density maps of the stellar distribution of the central galaxies that form in the GM-early (left column), Organic (centre column) and GM-late (right column) haloes. The galaxy shown in each case is that in the simulation from each assembly history family that adopts the same random number seed as the EAGLE simulations. The overlaid contours show the star-forming gas distribution, enclosing hydrogen column densities of $N_{\rm H}=10^{20}$, $10^{21}$, and $10^{22}$ cm$^{-2}$ in progressively darker shades of green. The field of view (and projected depth) of the maps is 80 ckpc. The specific star formation rate (sSFR) and stellar co-rotational kinetic energy fraction ($\kappa_{\rm co}$) are quoted for each galaxy in the lower row.}
\label{fig:galaxypic}
\end{figure*}

To illustrate the transformative effects of a modified assembly history on the properties of the central galaxy, we show in Fig.~\ref{fig:galaxypic} face-on and edge-on images of a present-day galaxy from each family. We show the galaxy from each assembly history family that adopts the same random number seed used by the EAGLE suite, but as noted in Section \ref{sec:methods:seeds}, this choice is equivalent to any other (i.e. the example was not ``cherry-picked'' and is representative of the overall population). The images are surface density maps of the stellar distribution with a field of view, and depth in the projection axis, of $80\ckpc$. The overlaid green contours show the distribution of the ISM within the same volume; the three progressively darker contours enclose hydrogen column densities of $N_{\rm H}=10^{20}$, $10^{21}$, and $10^{22}$ cm$^{-2}$ respectively. 

The Organic and GM-late assembly histories yield actively star-forming disc galaxies, with the GM-late galaxy exhibiting greater rotational support ($\kappa_{\rm co} = 0.50$) than the Organic case ($\kappa_{\rm co}=0.43$). The GM-early galaxy is quenched and exhibits a slightly oblate spheroidal morphology ($\kappa_{\rm co}=0.24$). The Organic and GM-late galaxies host extended discs of star-forming gas, sustaining sSFRs of $10^{-10.0}\peryr$ and $10^{-10.6}\peryr$ respectively; the GM-late galaxy hosts significantly more star-forming gas (quantified by the ISM mass fraction, $f_{\rm ISM} \equiv M_{\rm ISM}/M_\star=0.24$) than the Organic galaxy ($f_{\rm ISM} = 0.08$). In contrast, the GM-early galaxy is devoid of star-forming gas, with an instantaneous sSFR of zero; the value quoted in Fig.~\ref{fig:galaxypic} is integrated over the preceding 300 Myr for consistency with the results shown in Fig.~\ref{fig:ev:ssfr}.

In comparing the columns of Fig. \ref{fig:galaxypic}, one is comparing three versions of the same galaxy, in a halo of near-identical mass, embedded within the same large-scale environment. As we will demonstrate in the remainder of this section, the differences between these realisations lie in how significantly the content, density and cooling time of their CGM has been affected by AGN feedback, which must ultimately be determined by the halo assembly history as this is the only variable we adjust. 

\subsection{BH growth and AGN feedback}
\label{sec:results:eco:BH}

We begin investigating the cause of the differences shown in Section \ref{sec:results:eco:gal} by examining the influence of halo assembly history on the growth of the central BH. Previous studies indicate that these processes are fundamentally linked \citep{boothschaye10,boothschaye11} and, as shown in Fig. \ref{fig:ev:ssfr}, the quenching of star formation as a result of earlier halo assembly requires the injection of energy by AGN feedback. D20 identified that the correlations between $f_{\rm CGM}$ and the properties of galaxies (such as the sSFR and its morphology), and between $f_{\rm CGM}$ and proxies for the halo assembly history, are mediated by the expulsion of circumgalactic gas due to efficient AGN feedback. In both EAGLE and TNG, these effects are manifest in haloes that are sufficiently massive to host central BHs capable of delivering a quantity of feedback energy to the galaxy-CGM ecosystem that is comparable to the binding energy of the CGM gas. In EAGLE this corresponds to haloes with $M_{200} \gtrsim M_{\rm crit}$, i.e. those for which the buoyant transport of outflows heated by stellar feedback ceases to be efficient, resulting in the establishment of a quasi-hydrostatic hot halo. In TNG it corresponds to the haloes that host massive BHs ($M_{\rm BH} \gtrsim 10^8\Msun$), since AGN feedback is typically delivered in the efficient kinetic mode for such haloes in that model \citep{weinberger18}. 

\begin{figure}
\includegraphics[width=\columnwidth]{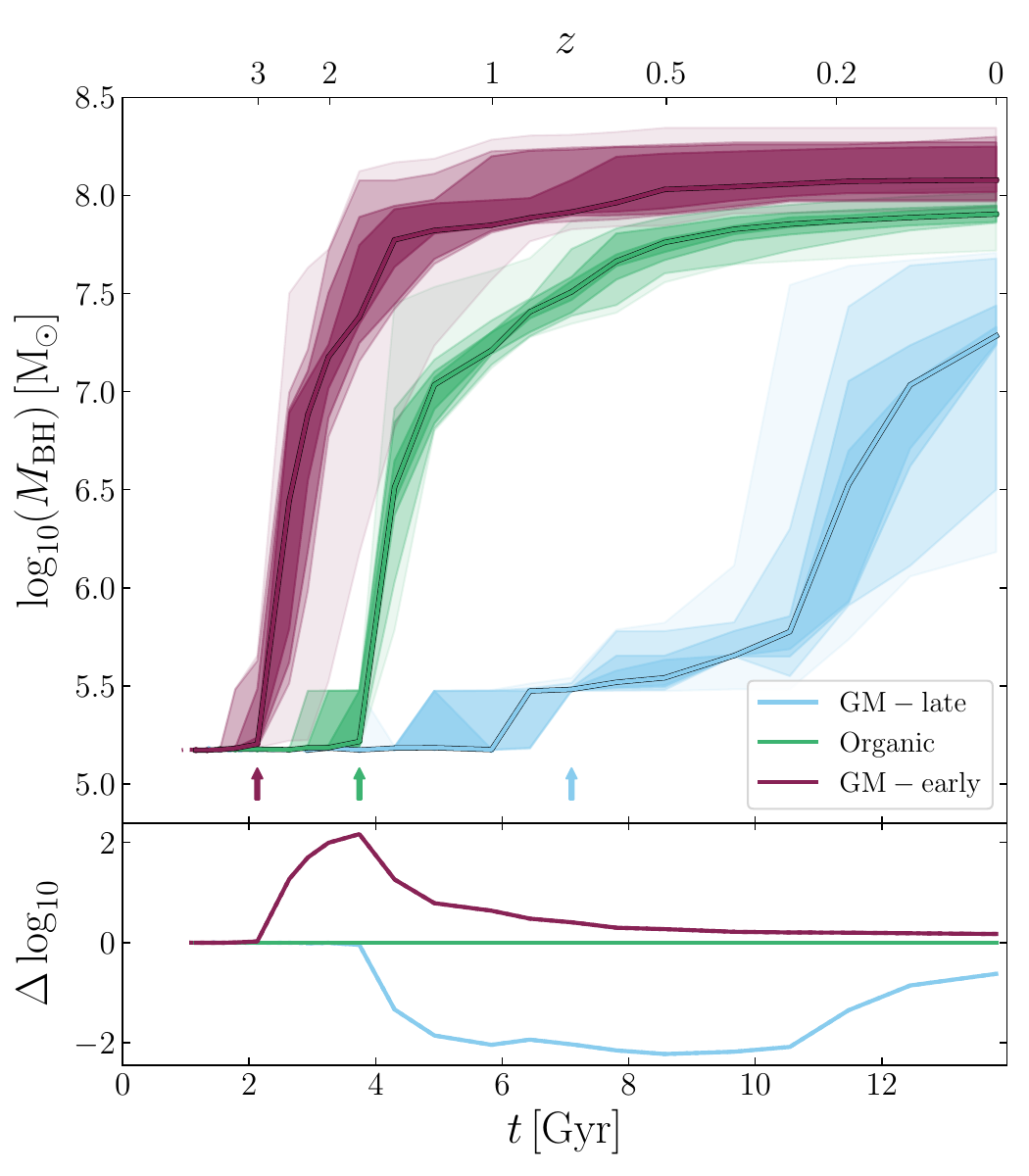}
\caption[The evolution of $M_{\rm BH}$ for the three families of simulations.]{The evolution of the BH mass, $M_{\rm BH}(t)$, for the three families of simulations, shown in the same fashion as Fig. \ref{fig:ev:ssfr}. Sub-panels show the deviation, $\Delta(t)$, of the logarithm of the median BH mass from that of the Organic case. Earlier halo assembly leads to the formation of a more massive central BH, and by extension the injection of more AGN feedback energy.}
\label{fig:ev:MBH}
 \end{figure}

Fig.~\ref{fig:ev:MBH} shows the evolution of the mass of the central BH\footnote{We define the central BH as the most massive BH particle in the system.}, $M_{\rm BH}(t)$, for the three families of simulations. Comparison of the median curves reveals that the central BHs in the GM-early family reach a greater present-day mass ($\log_{10}(M_{\rm BH}/\Msun)=8.08$, IQR$ = 0.30$ dex) than is the case for the Organic family ($\log_{10}(M_{\rm BH}/\Msun)=7.90$, IQR$ = 0.08$ dex), having commenced their rapid growth phases earlier. By $z=1$, the median $M_{\rm BH}$ of the GM-early family has reached 59 percent of its final value, while the median for the Organic case has reached only 20 percent of its final value. In marked contrast to these families, BHs in the GM-late family of simulations remain close to the seed mass until $z \simeq 1$, and do not grow rapidly until $z\simeq 0.3$. In consequence, they attain a significantly lower present-day mass ($\log_{10}(M_{\rm BH}/\Msun)=7.28$, IQR$= 0.20$ dex). The shaded regions in Fig.~\ref{fig:ev:MBH} indicate that the growth histories of individual BHs can vary due to the stochastic nature of EAGLE's feedback scheme; however there is clear separation between the three families of simulations, with overlap between only the most extreme cases. 

The early collapse of dark matter haloes (of a fixed present-day mass) leads to a higher concentration \citep[e.g.][]{neto07} and hence central binding energy. This is the case in our simulations; the GM-early system exhibits the highest intrinsic concentration, $c_{\rm DMO}=9.1$, followed by the Organic ($c_{\rm DMO}=6.6$) and GM-late ($c_{\rm DMO}=5.7$) systems. If a central BH is to self-regulate its own growth, one might expect that the ratio of the energy it injects through feedback to the binding energy of the baryons should, in the absence of other contributing influences, asymptote to a similar value regardless of assembly time. However, D20 showed that the ratio of the injected feedback energy to the (intrinsic) binding energy of the halo correlates negatively with assembly time in both EAGLE and TNG, such that the additional energy injected by the central galaxy of early-forming haloes ``overshoots'' the additional binding energy resulting from their higher concentration.

\begin{figure}
\includegraphics[width=\columnwidth]{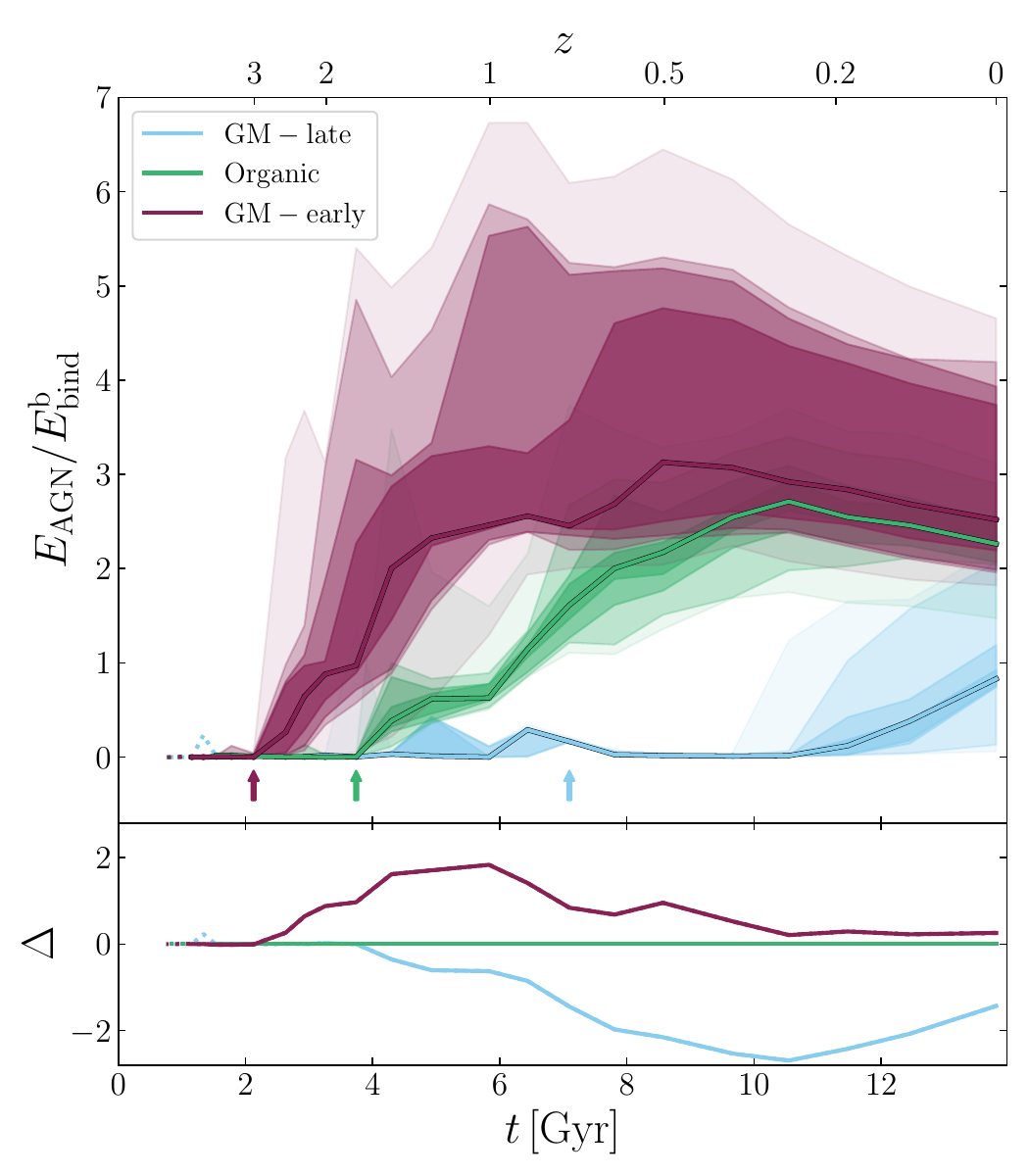}
\caption{The evolution of the $E_{\rm AGN}/E_{\rm bind}^{\rm b}$ ratio for the three families of simulations, shown in the same fashion as Fig. \ref{fig:ev:ssfr}. Here $E_{\rm AGN}$ is the total energy injected by AGN feedback and $E_{\rm bind}^{\rm b}$ is the intrinsic binding energy of the halo's baryons. Sub-panels show the deviation, $\Delta(t)$, of the median value from that of the Organic case. Earlier halo assembly leads to the injection of more AGN feedback energy relative to the intrinsic binding energy of the halo baryons.}
\label{fig:ev:agn_ov_bind}
\end{figure}

We therefore examine whether this ratio changes systematically in response to adjustment of the halo assembly history. Fig.~\ref{fig:ev:agn_ov_bind} shows the evolution of the $E_{\rm AGN}(z)/E_{\rm bind}^{\rm b}(z)$ ratio, where $E_{\rm AGN}(z)$ is the total energy injected by AGN feedback (defined per Equation ~\ref{eq:e_agn}). We focus only on the energy injected by AGN, since it is this mechanism that is principally responsible for circumgalactic gas expulsion in haloes of $M_{200} \gtrsim 10^{12.5}$ M$_{\odot}$ in EAGLE. $E_{\rm bind}^{\rm b}(z)$ is the intrinsic binding energy of the halo baryons, computed from the particle distribution of the evolving haloes in their counterpart DMONLY simulations. This self-consistently accounts for differences in the structure of the haloes induced by their mass accretion histories (see Section~\ref{sec:methods:cooling}). 

Prior to the onset of the efficient growth of the BH, $E_{\rm AGN} \ll E_{\rm bind}^{\rm b}$. Once $M_{200}\simeq M_{\rm crit}$, the rapid growth of the BH results in a rapid increase of $E_{\rm AGN}$, such that the ratio $E_{\rm AGN} / E_{\rm bind}^{\rm b}$ stabilises at a value of order unity. In general, the ratio settles at a value greater than unity because radiative cooling inhibits the unbinding of circumgalactic gas, and because a fraction of the gas unbound at early times can re-accrete later as the halo potential grows \citep[see e.g.][]{mitchell20}. The halo reaches $M_{\rm crit}$ at very different times for the three assembly histories, and the increase in $E_{\rm AGN}/E_{\rm bind}^{\rm b}$ follows suit. The final median ratio for the GM-early case is highest $(E_{\rm AGN}/E_{\rm bind}^{\rm b}=2.52$, IQR$=1.94$), followed by the Organic case (2.26, IQR$=0.44$), while the ratio for the GM-late case is far lower (0.83, IQR$=0.45$). The scatter in the ratio is equivalent to the scatter in $M_{\rm BH}$ (since $E_{\rm bind}^{\rm b}$ is computed from a DMONLY run and is independent of the chosen seed) and is greatest for the GM-early case. 

The assembly history of the halo therefore appears to directly influence how much energy, beyond that required to unbind the halo's baryons, is injected into the galaxy-CGM ecosystem by the central BH. This connection was also identified by D20 through examination of statistical correlations in the EAGLE Ref-L100N1504 simulation. Critically, however, the differences in $M_{\rm BH}$ and $E_{\rm AGN} / E_{\rm bind}^{\rm b}$ between the Organic, GM-early and GM-late haloes can be {\it exclusively} linked to differences in their assembly histories in our controlled experiment. This is not the case for statistical analyses of populations drawn from large volumes, where factors such as environment may also play a role.

\subsection{CGM mass fraction}
\label{sec:results:eco:CGM}

\begin{figure}
\includegraphics[width=\columnwidth]{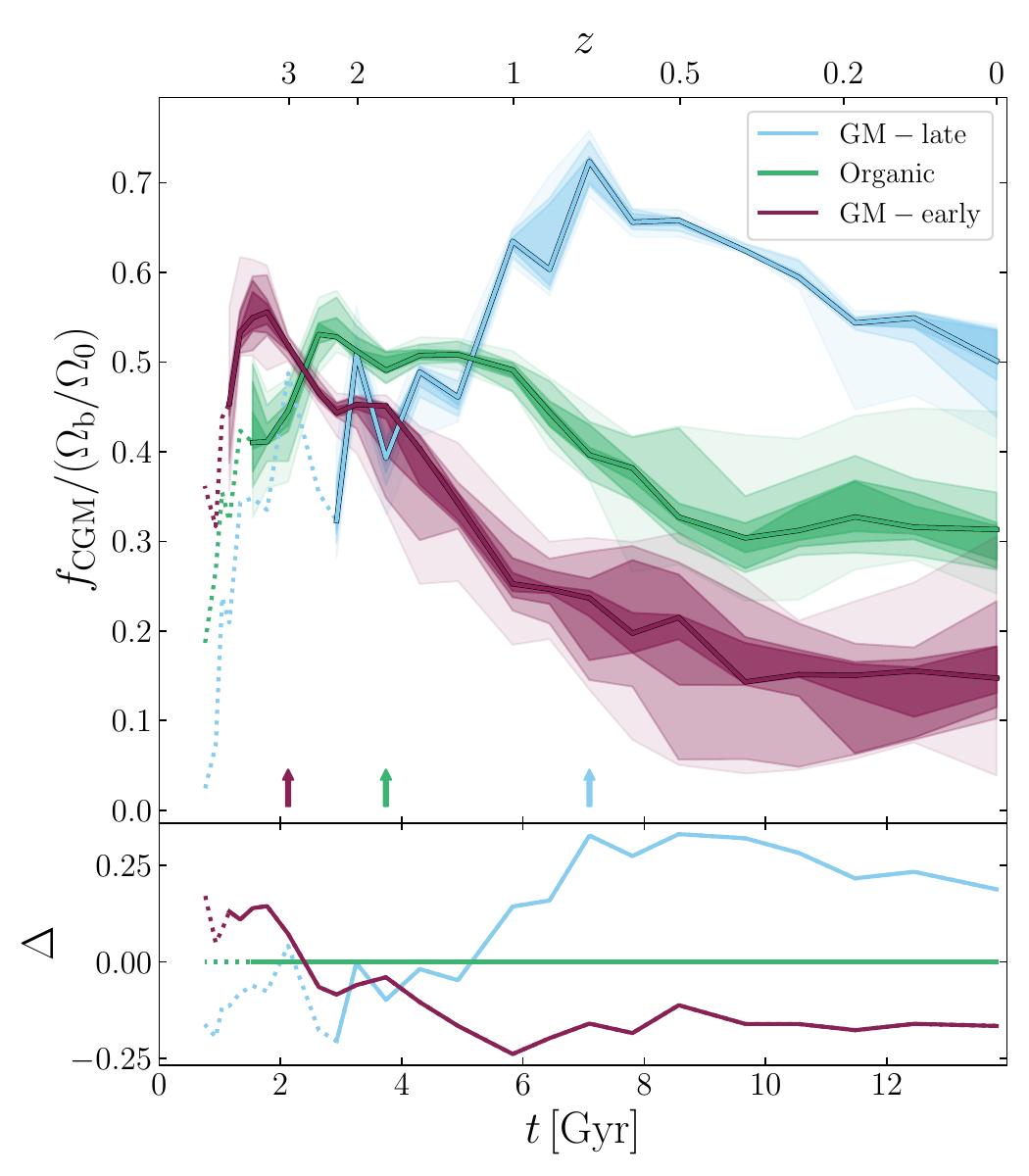}
\caption[The evolution of $f_{\rm CGM}$ for the three families of simulations.]{The evolution of the CGM mass fraction, $f_{\rm CGM}(t)\equiv M_{\rm CGM}(t)/M_{200}(t)$, normalised by the cosmic average baryon fraction, $\Omega_{\rm b}/\Omega_0$, for the three families of simulations, shown in the same fashion as Fig. \ref{fig:ev:ssfr}. Sub-panels show the deviation, $\Delta(t)$, of the median value from that of the Organic case. Earlier halo assembly leads to a more significant depletion of baryons from the CGM.}
\label{fig:ev:fCGM}
\end{figure}

We now turn to the effects of these differences in AGN feedback on the baryon content of the CGM. Fig.~\ref{fig:ev:fCGM} shows the evolution of the CGM mass fraction, $f_{\rm CGM}(t)$, normalised by the cosmic baryon fraction, $\Omega_{\rm b}/\Omega_{\rm 0}$, for our three families of simulations. The halo in the GM-late family of simulations has the highest CGM mass fraction at the present day, $f_{\rm CGM}/(\Omega_{\rm b}/\Omega_0)=0.50$ (IQR$=0.06$), followed by the Organic case (0.31, IQR$=0.05$), and the GM-early case (0.15, IQR$=0.07$). 
 
In all three families, $f_{\rm CGM}$ is low at early epochs, likely because the potential of the nascent halo is too shallow to accrete photoionised gas after the epoch of reionisation. As the halo grows, $f_{\rm CGM}$ quickly increases to a peak, ${\rm max}(f_{\rm CGM}) = (0.62,0.58,0.76)$ for GM-early, Organic and GM-late, respectively, but then begins to decline towards its $z=0$ value over several Gyr, with the decline broadly commencing when the halo mass reaches $M_{\rm crit}$ (denoted by the coloured arrows). Comparison of the tracks in Fig.~\ref{fig:ev:fCGM} with those of Fig.~\ref{fig:ev:MBH} illustrates that, as per the findings of \citet{opp20a}, a strong decline in $f_{\rm CGM}$ generally follows shortly after periods of rapid growth of the central BH, which is coincident with the halo reaching $M_{\rm crit}$ \citep{bower17}. AGN feedback acts to expel baryons from the halo, but has also been shown to reduce the baryon fraction of material infalling onto EAGLE haloes \citep{wright20}, further suppressing $f_{\rm CGM}$ through the prevention of accretion.

As noted in Section~\ref{sec:results:mod_histories}, the modified haloes reach $M_{\rm crit}$ significantly earlier (GM-early) and later (GM-late) than the Organic case, resulting in a markedly different evolution of $f_{\rm CGM}$. By $z=1$ the GM-early halo has already been strongly depleted of circumgalactic gas, $f_{\rm CGM}/(\Omega_{\rm b}/\Omega_0)=0.25$, IQR$=0.04$, while the Organic halo (0.49, IQR$=0.02$) is only slowly being depleted of baryons prior to a more rapid depletion at $z<1$, and the GM-late halo (0.63, IQR$=0.01$) does not begin to be be depleted until $z\lesssim 0.7$. The strong correlation between the halo assembly history and the present-day value of $f_{\rm CGM}$ from large galaxy samples in EAGLE and TNG (seen in D19 and D20) is therefore reproduced here in direct response to systematic adjustment of the assembly history of an individual halo.

\begin{figure}
\includegraphics[width=\columnwidth]{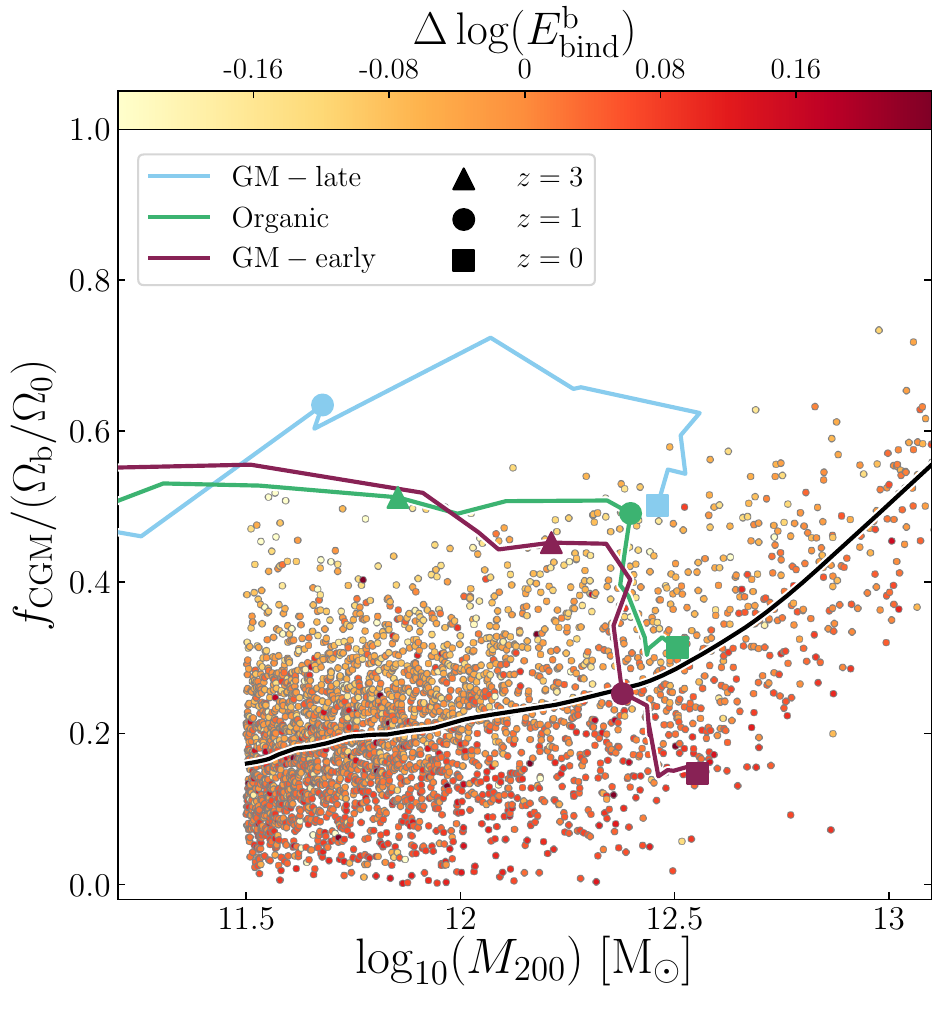}
\caption{Present-day CGM mass fractions, $f_{\rm CGM}(t)\equiv M_{\rm CGM}(t)/M_{200}(t)$, of haloes in the EAGLE Ref-L100N1504 simulation, normalised by the cosmic average baryon fraction, $\Omega_{\rm b}/\Omega_0$, as a function of halo mass, $M_{200}$, with the solid curve denoting the running median. Symbols are coloured by residuals about the relationship between the intrinsic binding energy of the inner halo, $E_{\rm DMO}^{2500}$, and $M_{200}$. The evolution of the median $f_{\rm CGM}/(\Omega_{\rm b}/\Omega_0)$ with the median $M_{200}$ for the three assembly histories is overlaid, with their locations on the plot at $z=0$, $z=1$ and $z=3$ denoted by squares, circles and triangles respectively. At $z=0$, the Organic system lies close to the median relation, while the systems realised with modified initial conditions span the scatter in the Ref-L100N1504 simulation.}
\label{fig:scatter:fCGM}
\end{figure}

Fig.~\ref{fig:scatter:fCGM} shows the present-day $f_{\rm CGM}-M_{200}$ relation of haloes of $M_{200}>10^{11.5} \Msun$ in the EAGLE Ref-L100N1504 simulation as a cloud of coloured points, with the running median relation obtained with the LOWESS algorithm \citep{cleveland79} shown as a black line. Symbols are coloured by the residuals about the LOWESS running median $\log_{10}(E_{\rm bind}^{\rm b})$ as a function of $M_{200}$. As detailed in Section \ref{sec:methods:cooling}, $E_{\rm bind}^{\rm b}$ represents the binding energy of the underlying dark matter structure (rescaled by $\Omega_{\rm b}/\Omega_0$), and is thus an effective proxy for the halo assembly time that is simple to compute without the need to examine merger trees; a greater binding energy corresponds to an earlier assembly time. The overlaid tracks show the evolution of (the median) $f_{\rm CGM}-M_{200}$ relation of the GM-early, Organic and GM-late simulations; values at $z=3$, $z=1$ and $z=0$ are denoted by large triangle, circle and square symbols, respectively. At the present day, the CGM mass fraction of the Organic halo is similar to the running median value for the EAGLE population. The halo was not explicitly selected on this criterion, however the selection of an unquenched present-day galaxy with a typical star formation rate makes it unlikely that the Organic case would deviate far from the median relation in EAGLE. 

The haloes with adjusted assembly histories yield present-day circumgalactic gas fractions that reside towards the extremities of the scatter for EAGLE haloes of $M_{200}\approx 10^{12.5}$ M$_\odot$. If one compares with EAGLE haloes in a $0.2 \dex$ wide bin centred on this mass, the GM-late halo represents a $+2.0\sigma$ shift from the median in terms of $f_{\rm CGM}$, while the GM-early halo represents a $-1.3\sigma$ shift. Clearly, if these systems were to occur ``organically'' in the simulation, they would be amongst the most CGM-rich (GM-late) and CGM-poor (GM-early) haloes of their mass. In terms of $E_{\rm bind}^{\rm b}$, the GM-late halo represents a $-1.5\sigma$ shift from the median, while the GM-early halo represents a $+1.2\sigma$ shift; as is clear from the symbol colouring, the GM-late (GM-early) cases yield $f_{\rm CGM}$ values similar to haloes in the EAGLE simulation with low (high) values of $E_{\rm bind}^{\rm b}$. 


The evolution of the $E_{\rm AGN}/E_{\rm bind}^{\rm b}$ ratio, shown in Fig. \ref{fig:ev:agn_ov_bind}, provides an intuitive explanation for the evolution of the CGM mass fraction. At $z=1$, the energy injected via AGN feedback has already exceeded the binding energy of the baryons in the GM-early case ($E_{\rm AGN}/E_{\rm bind}^{\rm b} \simeq 2.5$, IQR$=3.2$), and consequently the CGM of the halo in the GM-early family has already been depleted of a significant fraction of its mass. In the Organic case the two energies are comparable ($E_{\rm AGN}/E_{\rm bind}^{\rm b} \simeq 0.6$, IQR$=0.2$) and the CGM is about to be depleted, while in the GM-late case the injected energy remains much less than the intrinsic binding energy ($E_{\rm AGN}/E_{\rm bind}^{\rm b}\sim 10^{-3}$, IQR$=0.1$) and the CGM remains gas-rich. 

It is plausible that there is a close coupling of this ratio to the CGM mass fraction, such that the injection of more energy relative to the binding energy of the baryons yields a lower $f_{\rm CGM}$, and this is largely supported by comparison of Figs. \ref{fig:ev:agn_ov_bind} and \ref{fig:ev:fCGM}. There is, however, substantial overlap between the scatter in the $E_{\rm AGN}/E_{\rm bind}^{\rm b}$ ratio for the GM-early and Organic simulations, and they reach similar final $E_{\rm AGN}/E_{\rm bind}$ despite exhibiting significantly different $f_{\rm CGM}$ (see Fig. \ref{fig:ev:fCGM}). A potential explanation for this is that earlier energy injection is more efficient at evacuating the CGM; gas in the vicinity of the BH is heated by a fixed temperature increment in EAGLE's feedback scheme, and therefore reaches a higher fraction (or multiple) of the halo virial temperature (or entropy) if heated whilst in a shallower potential.

The results presented in this section, and in Section \ref{sec:results:eco:BH}, provide new and complementary evidence supporting the hypotheses of D19 and D20: in the galaxy populations of EAGLE and TNG, the binding energy of the underlying dark matter halo (a good proxy for assembly time) correlates positively with $M_{\rm BH}$ and $E_{\rm AGN}/E_{\rm bind}^{\rm b}$, which both in turn correlate negatively with $f_{\rm CGM}$, thus connecting the assembly histories of dark matter haloes with their circumgalactic mass fractions. We can now confirm a direct and causal connection between these processes, as the only difference between our three families of simulations lies in the halo accretion history. 


\subsection{Structure and thermodynamic state of the CGM}
\label{sec:thermo}

\begin{figure*}
\includegraphics[width=\textwidth]{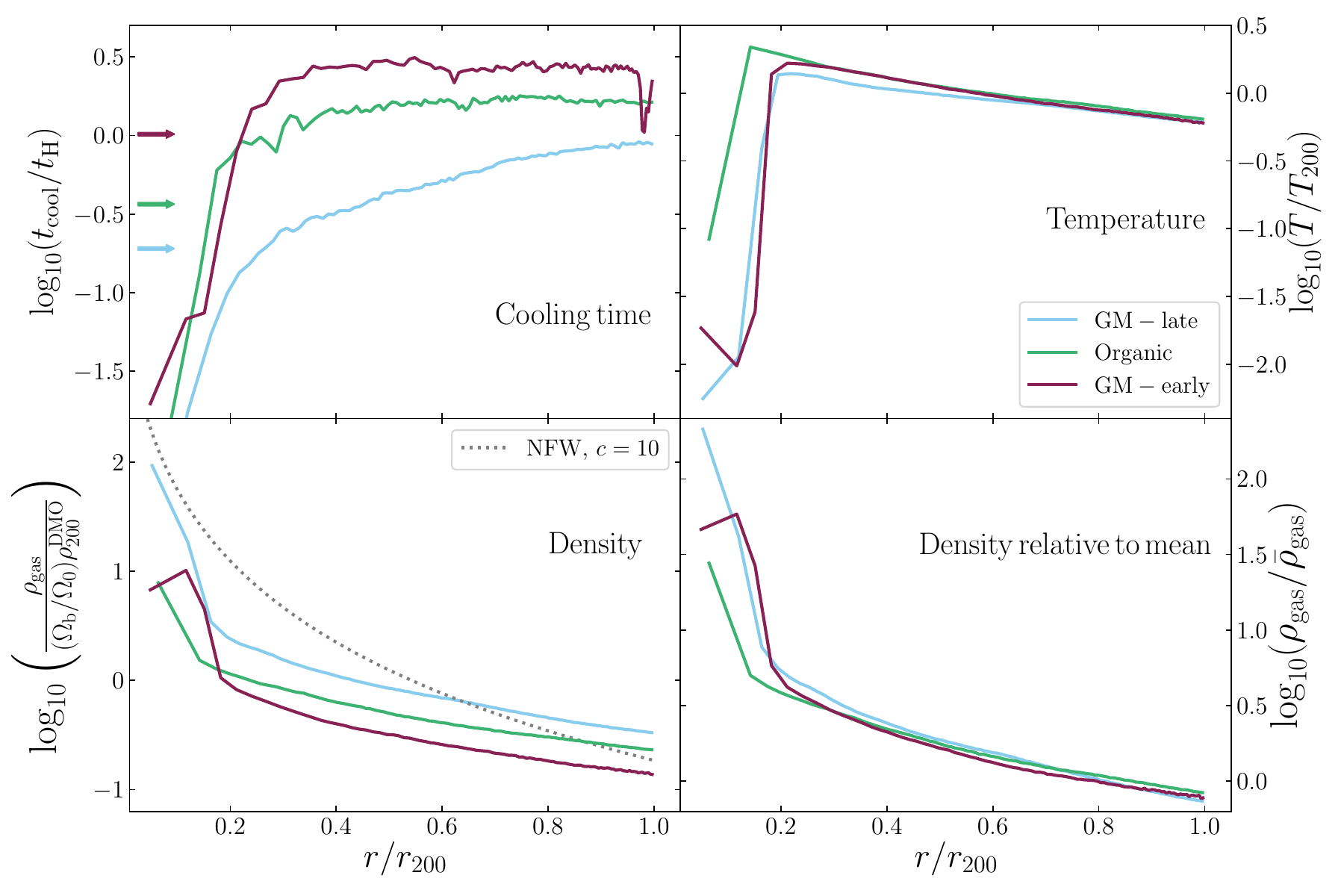}
\caption{Radial profiles of the cooling time ($t_{\rm cool}$, calculated as described in Section \ref{sec:methods:cooling}), median temperature ($T$) and median density ($\rho_{\rm gas}$) of fluid elements comprising the CGM, for the three assembly history families. The solid curves show stacked values for particles from all nine simulations of each family. The cooling time is normalised to the Hubble time, $t_{\rm H}$, and the temperature is normalised to the virial temperature $T_{200}$. Density profiles are shown normalised to the mean gas density expected in the absence of radiative processes (lower left panel), and to the true mean density of the halo, $\bar{\rho}_{\rm gas}$ (lower right panel). An NFW profile with a concentration $c=10$ is shown as a dotted line in the lower-left panel. Earlier halo assembly leads to longer CGM cooling times primarily because the mean density of the CGM is reduced in response to AGN-driven gas expulsion.}
\label{fig:radial}
\end{figure*}

The results presented in the previous sections show that the expulsion of circumgalactic gas by AGN can be effectively ``dialled'' up or down via adjustment of the halo assembly history. We turn now to the effects of these differences on the structure and physical state of the CGM, and to the consequences for the central galaxy.

D20 showed that in both EAGLE and TNG, haloes that are gas-poor exhibit longer characteristic cooling times (and characteristic entropies) than gas-rich haloes at fixed mass. The radiative cooling time distributions of fluid elements in gas-poor haloes were also shown to be systematically greater than those of fluid elements in gas-rich haloes of similar mass. The effect of this shift is that the CGM is less able to cool and replenish the ISM when the latter is depleted by star formation and feedback processes, ultimately facilitating quenching and morphological transformation. We therefore now examine the CGM cooling times of our genetically-modified haloes with the aim of establishing a causal link between early halo assembly and the suppression of cooling from the CGM at late times.

In Fig. \ref{fig:radial} we show present-day radial profiles of several properties of the circumgalactic gas within $r_{200}$ in our three families of simulations. To obtain these curves for each family, particles in the CGM realised with all nine random number seeds are stacked\footnote{Since all nine simulations in each family are equivalent, stacking them affords affords superior particle sampling of the radial profiles `for free'.} as a function of their radial distance, $r$, from the halo centre in 100 bins of equal particle number. The upper left panel shows radial profiles of the CGM cooling time, $t_{\rm cool}$, normalised by the Hubble time, $t_{\rm H}$. As described in Section \ref{sec:methods:cooling}, we define $t_{\rm cool}$ as the sum of the internal energies of the particles in each bin divided by the sum of their bolometric luminosities, in order to mimic an observational measurement. In all three cases, $t_{\rm cool}$ rises monotonically as a function of $r/r_{200}$ and indicates the presence of some efficiently-cooling gas ($t_{\rm cool}\ll t_{\rm H}$) in the centres of all three families, with the bulk of the gas exhibiting long cooling times for $r \gtrsim 0.2r_{200}$. Gas in the GM-early family exhibits the longest characteristic $t_{\rm cool}$, followed by the Organic and GM-late families. We quantify these differences with the characteristic cooling time at $r=0.5r_{200}$; $t_{\rm cool}(0.5r_{200})=2.9t_{\rm H}$ for the GM-early family, 1.6$t_{\rm H}$ for the Organic family and 0.4$t_{\rm H}$ for the GM-late family. D20 computed global $t_{\rm cool}$ values for haloes in EAGLE and TNG by performing the same calculation considering all particles within $r_{200}$. We denote these values, $t_{\rm cool}^{\rm halo}$, for our three families of simulations with arrows in the upper left panel of Fig. \ref{fig:radial}; $t_{\rm cool}^{\rm halo}=14.1$, 5.0 and 2.6 Gyr for the GM-early, Organic and GM-late families respectively. In all three cases, $t_{\rm cool}^{\rm halo}<t_{\rm cool}(0.5r_{200})$ since the former is strongly weighted to the most rapidly-cooling material in the halo. Differences in the expulsion of baryons from the CGM clearly induce strong changes in the ability of the CGM to cool efficiently, replenish the ISM and sustain star formation.

To elucidate the cause of these shifts in the cooling time profiles, we show radial profiles of the median gas temperature and density in the upper-right and lower-left panels of Fig. \ref{fig:radial} respectively. The temperature profiles are normalised to the halo virial temperature, $T_{200}\equiv G M_{200}\mu m_{\rm p}/2k_{\rm B}r_{200}$, where $G$ is Newton's gravitational constant, $m_{\rm p}$ is the proton mass, $k_{\rm B}$ is the Boltzmann constant and we assume a mean molecular weight, $\mu$, of 0.59, consistent with a fully ionised primordial gas. The temperature profiles indicate the presence of cool gas in the centres of all three families, a rapid increase in temperature with increasing radius for $r<0.2r_{200}$, followed by a steady decline in temperature towards $r=r_{200}$. The temperature profiles for the three simulation families are near-identical throughout the bulk of the halo; $T(0.5r_{200})=1.09T_{200},\,1.11T_{200}$ and $0.96T_{200}$ for the GM-early, Organic and GM-late cases respectively. It appears that the bulk of the CGM is quasi-hydrostatic ($T\sim T_{200}$), and that differences in the expulsion of the CGM do not induce strong changes in its characteristic temperature. Circumgalactic gas heated by AGN feedback must therefore leave the halo quickly without strongly heating the gas that remains.

The density profiles in the lower-left panel of Fig. \ref{fig:radial} show the median gas density, $\rho_{\rm gas}$, normalised to the mean density of the halo baryons expected in the absence of any expulsion, $\rho_{200}^{\rm cosmic}=(\Omega_{\rm b}/\Omega_0)\rho_{200}^{\rm DMO}$, where $\rho_{200}^{\rm DMO}$ is the mean density of the halo's counterpart in a purely gravitational DMONLY simulation. A Navarro-Frenk-White (NFW) profile with a concentration, $c$, of 10, is shown with a dotted line to illustrate the profile expected in the absence of dissipative or expulsive processes. The median density profiles of our three simulation families fall well below the NFW profile as a result of baryon expulsion \citep[see also][]{crain10,kelly20}; critically, earlier assembly leads to lower gas densities in the CGM. Over the majority of the galactocentric radius, the three families exhibit profiles of similar shape, but the GM-early family exhibits the lowest normalisation $(\rho_{\rm gas}(0.5r_{200})=0.32\rho_{200}^{\rm cosmic})$, followed by the Organic $(\rho_{\rm gas}(0.5r_{200})=0.49\rho_{200}^{\rm cosmic})$ and GM-late $(\rho_{\rm gas}(0.5r_{200})=0.84\rho_{200}^{\rm cosmic})$ families.

The lower-right panel of Fig. \ref{fig:radial} shows the median density profiles normalised to $\bar{\rho}_{\rm gas}$, the mean density of gas within $r_{200}$. When normalised in this fashion, the density profiles of the three families are very similar, revealing that differences in the fraction of the halo baryons expelled by feedback do not strongly affect the form of the CGM density profile, only its overall mass and mean density. This suggests that the outflows driven by feedback are able to entrain and expel gas from the halo at all radii, and/or that halo quickly reconfigures itself at a lower density following expulsive feedback episodes.

We conclude that AGN feedback elevates the cooling time of the CGM primarily as a consequence of the reduction of its characteristic density. Since the radiative cooling rate of gas is proportional to the square of its density, this change extends the characteristic cooling time of the CGM and inhibits the efficient replenishment of the ISM, leading to sustained quenching of star formation. Our controlled experiment reveals that the magnitude of this effect is greater for haloes that assemble earlier and experience more AGN feedback over cosmic time, thus explaining the marked differences in the evolution of the sSFR for the three families of simulations shown in Fig. \ref{fig:ev:ssfr}.

These differences in the cooling properties of the CGM are likely also the cause of the differences in morphological evolution between our central galaxies. D20 speculated that the connection between assembly history and morphology arises because the elevation of the CGM cooling time (in response to circumgalactic gas expulsion) inhibits the replenishment of the interstellar gas in galaxy discs, which would otherwise stabilise them against transformation by mergers, tidal interactions and gravitational instabilities, and enable re-growth of the stellar disc \citep[see e.g.][]{robertson06,hopkins09,font17}. This hypothesis is borne out by our results; as shown in Fig. \ref{fig:ev:kappa}, the transformation in the Organic and GM-early cases is gradual and proceeds over several gigayears, contrary to the rapid transformation one would expect from a major merger.

\section{Summary and Discussion}
\label{sec:paper3:summary}

We have investigated the impact of the assembly history of a dark matter halo on the properties of its central galaxy and circumgalactic medium (CGM), using a suite of zoom simulations with controlled assembly histories. We have used the ``genetic modification'' (GM) technique \citep[][see also \citealt{rey18}]{roth16,pontzen17} to adjust the assembly history of the halo, whilst ensuring that its present day mass ($M_{200}$) is not significantly altered. 

This study was motivated by the previous identification of several correlations at fixed halo mass in the EAGLE and IllustrisTNG (TNG) cosmological, hydrodynamical simulations of galaxy formation, which suggested that differences in assembly history drive the scatter in the CGM mass fractions of dark matter haloes, mediated by differences in the integrated feedback from active galactic nuclei (AGN) injected into the system. These correlations, presented in \citet[][D19]{davies19} and \citet[][D20]{davies20}, also indicate that differences in the expulsion of CGM baryons by AGN feedback play a key role in governing the properties of $\sim L^\star$ central galaxies, particularly with regard to their star formation history and morphological evolution. The quenching and morphological transformation of $\sim L^\star$ galaxies therefore appears to be directly linked to the assembly history of their host dark matter haloes, which is determined only by the underlying cosmogony. 

A limitation of these findings, however, is that they were arrived at by comparing {\it different} haloes in cosmologically-representative volumes, thus precluding the exclusion of other possible influences, such as environment. Here, we have moved beyond a purely statistical analysis, and established a causal and exclusive link between these processes via the use of a controlled numerical experiment. We use zoom simulations of the {\it same} halo, creating three families of simulations with different assembly histories. Beside the unmodified ``Organic'' case, we have created complementary initial conditions that shift the halo assembly history to earlier (``GM-early'') and later (``GM-late'') times. 

The galaxy we have studied was drawn from a simulation of a periodic volume evolved with the Reference EAGLE model \citep{schaye15,crain15}. It was selected to be a present-day moderately star-forming (sSFR$=10^{-10.2}\peryr$) central galaxy of stellar mass $M_{\star,{\rm 30kpc}}=4.3\times 10^{10}$ M$_\odot$ hosted by a halo of mass $M_{200}=3.4\times 10^{12}$ M$_\odot$, chosen to match the halo mass scale at which the correlations between the CGM mass fraction and halo assembly are the strongest in EAGLE and TNG. The zoom simulations were carried out with the `Recal' EAGLE model \citep{schaye15}; we also use counterpart simulations with only collisionless gravitational dynamics (DMONLY) and full-physics simulations where no black holes are seeded and no AGN feedback occurs (NOAGN). To quantify the effects of the stochasticity inherent to EAGLE's subgrid treatments of star formation and feedback, we evolve the simulation from the initial conditions with nine different initial seed values for the quasi-random number generator used by these routines. We therefore quote the emergent properties of the halo and galaxy at a given epoch as the median value of the property measured for all nine simulations, and quantify the scatter with the interquartile range (IQR).

Our results can be summarised as follows:

\begin{enumerate}

    \item The three families of assembly histories yield haloes with very similar final halo masses; the GM-early and GM-late cases differ from the Organic case by only 0.04 dex. The families exhibit strong differences in their assembly histories, as intended, a result of the genetic modifications. At $z=2$, the epoch at which the overdensity of the matter comprising the halo is specified by the GM technique, the GM-early system has already assembled 47\% of its final mass, while the Organic and GM-late systems have assembled 23\% and 3\% of their final masses respectively (Fig.~\ref{fig:ev:m200}).
    
    \item The halo assembly history has a marked influence on the star formation history of the central galaxy. All realisations of the GM-late and Organic haloes remain actively star-forming at the present day (i.e. sSFR$ >10^{-11}\peryr$), while the GM-early system is quenched at the present day in six of the nine realisations. A systematic shift in the assembly history of the dark matter halo hosting a star-forming $\sim L^\star$ galaxy can therefore result in galaxy quenching. AGN feedback is crucial for mediating this connection; in its absence, all realisations of all three families of assembly histories yield galaxies that are star-forming at the present day (Fig.~\ref{fig:ev:ssfr}).
    
    \item The decline of the sSFR in the GM-early and Organic systems is accompanied by a decrease in the degree of rotational support in the stellar disc, quantified using the stellar co-rotational kinetic energy fraction, $\kappa_{\rm co}$. In the GM-early case, the central galaxy experiences a strong morphological evolution from disc-like to spheroidal, while in the Organic and GM-late cases the central galaxy remains disc-like. (Fig.~\ref{fig:ev:kappa}, see also the images in Fig.~\ref{fig:galaxypic}).
    
    \item In all three families, the onset of rapid BH growth is broadly coincident with $M_{200}(z)$ reaching the threshold, $M_{\rm crit}(z)$, at which EAGLE's stellar feedback is expected to cease efficiently regulating star formation \citep{bower17}. Prior studies using cosmological simulation indicate that earlier-forming and more tightly-bound haloes foster the growth of more massive BHs \citep{boothschaye10,boothschaye11}, and the GM-early system indeed hosts a more-massive central BH at the present day than the Organic system, which in turn hosts a more massive BH than the GM-late system (Figure \ref{fig:ev:MBH}). 
    
    \item Once $M_{200}$ exceeds $M_{\rm crit}(z)$, the total energy injected by the AGN, $E_{\rm AGN}$, increases quickly to become comparable with the intrinsic binding energy of the halo baryons, $E_{\rm bind}^{\rm b}$. The final ratio of these quantities is greatest for the GM-early case, followed by the Organic case, then the GM-late case (Fig. \ref{fig:ev:agn_ov_bind}). The differences in halo binding energy induced by the adjustment of the assembly history therefore modulate the total energy injected by AGN feedback relative to the binding energy.
    
    \item The CGM mass fraction, $f_{\rm CGM}$, declines for all three GM cases once the halo mass exceeds $M_{\rm crit}(z)$ and the BH begins to grow rapidly and efficiently inject energy as AGN feedback. The onset of baryon expulsion in each system is coincident with $E_{\rm AGN}\simeq E_{\rm bind}^{\rm b}$. In accord with the correlations found between $f_{\rm CGM}$ and proxies for the halo assembly time in EAGLE and TNG (D19,D20), the GM-late halo is the most gas-rich at the present day, followed by the Organic and GM-early haloes, demonstrating the strong influence of assembly time on the baryon content of the CGM (Fig.~\ref{fig:ev:fCGM}). The induced differences in $f_{\rm CGM}$ are comparable with the scatter in the present-day $f_{\rm CGM}-M_{200}$ relation of the galaxy population in the EAGLE Ref-L100N1504 simulation (Fig.~\ref{fig:scatter:fCGM}).
    
    \item The AGN-driven expulsion of baryons elevates the cooling time of the baryons remaining in the CGM. This occurs primarily because the expulsion reconfigures the CGM at a lower density. Since the radiative cooling rate of gas is proportional to the square of its density, the expulsion strongly influences the ability of the CGM to replenish the ISM as it is consumed by star formation and ejected by feedback process. (Fig. \ref{fig:radial}).

\end{enumerate}

Our findings demonstrate conclusively that, in the EAGLE model, present-day $\sim L^\star$ galaxies are significantly influenced by the assembly history of their host dark matter haloes. We identify a clear sequence of events: haloes that form earlier than is typical for their mass foster the growth of more massive  central BHs, which inject more feedback energy into their surrounding gas relative to the binding energy of the halo's baryons. This expels a greater fraction of the CGM, reconfiguring it at a lower mean density, elevating its cooling time and inhibiting replenishment of the ISM. This facilitates the quenching and morphological transformation of the central galaxy. While the results of D19 and D20 suggested this sequence indirectly, via the identification of correlations in large statistical samples, here we can have greater confidence that assembly history is the fundamental driver. This is because we consider the evolution of an individual halo, and thus minimise or eliminate the influence of other variables.

We have only considered the EAGLE simulation model in this study, however the results of D20 demonstrated that this picture is similarly applicable to the TNG model, signalling an important consensus between the two state-of-the-art models of the evolution of the galaxy population. If this controlled experiment were replicated using the TNG model we anticipate recovering qualitatively similar outcomes, but with differences in detail stemming from the use of very different subgrid treatments of the feedback processes associated with star formation and BH growth. As shown by D20, the onset of efficient baryon expulsion in EAGLE occurs when high BH accretion rates are reached, since the efficiency of AGN feedback is fixed; this happens at earlier times for earlier-assembling systems. In contrast, the onset of efficient expulsion in TNG occurs when the BH accretion rate is low compared to the Eddington rate, and the AGN injects feedback energy in the high-efficiency `kinetic' mode. This is typically the case once $M_{\rm BH}$ is greater than the ``pivot mass'' of $10^8$ M$_\odot$, and this threshold is likely reached at earlier times for central BHs hosted by earlier-assembling haloes. The triggering of efficient AGN feedback is key to the sequence of events described above, and it is difficult to conceive of a model in which this does not occur at earlier times for earlier-forming haloes, encouraging us to posit that our findings are likely general.

The differences in the star formation histories of our modified systems from that of the Organic system can be intuitively understood as the result of feedback-induced differences in ISM replenishment from the CGM. The origin of the differences in the evolution of the galaxy kinematics are less clear, however, as modifying the assembly history can potentially change the angular momentum of the baryons from which the $z=0$ stellar population forms through several routes. One one hand, linear theory dictates that the angular momentum of dark matter haloes scales as $a^{3/2}$ (where $a$ is the cosmological expansion factor) until it is ``frozen in" at the turnaround time, resulting in later-assembling haloes exhibiting higher halo spins at fixed $z=0$ mass \citep{white84,zavala08}. On the other hand, cosmological simulations suggest that the galaxy angular momentum is markedly affected by feedback processes, which preferentially expel baryons with lower angular momentum \citep{brook11}. The injection of more AGN feedback energy in earlier-assembling haloes may therefore increase the angular momentum of the condensed baryons through this selective removal. Other studies have demonstrated that galaxy kinematics are sensitive to how coherently aligned the angular momentum of inflowing material is with the extant galaxy, such that accretion from a quasi-hydrostatic gas corona provides coherent rotational support, while accretion through (potentially misaligned) cold flows and mergers leads to spheroidal morphologies \citep[][see also \citealt{aumer13}]{sales12}. Our results suggest that that the AGN-driven elevation of the CGM cooling time (through expulsion) is the dominant effect, as it inhibits further accretion from the CGM and leaves the rotational kinematics of the disc to be gradually eroded by non-coherent accretion of gas and ex-situ stars, and by instabilities and mergers \citep[e.g.][]{delucia11,clauwens18}. This mechanism explains the gradual decline in $\kappa_{\rm co}$ following the expulsion of circumgalactic gas in the GM-early and Organic systems (Fig. \ref{fig:ev:kappa}).

D19 demonstrated that the soft (0.5-2.0 keV) X-ray luminosity of the CGM is an attractive observational proxy for $f_{\rm CGM}$ (in the case of $\sim L^\star$ central galaxies), and that it correlates strongly with properties such as the BH mass and star formation rate in the same fashion as $f_{\rm CGM}$. Detailed study of the X-ray-luminous CGM of galaxies with diverse star formation rates, morphologies and central BH masses therefore presents a plausible means of testing the predictions advanced here, though such studies await the launch of next-generation X-ray observatories such as {\it Athena} \citep{barret16} and {\it Lynx} \citep{ozel18}. Prior to the advent of these missions, a promising alternative is to appeal to survey data from the {\it eROSITA} instrument aboard the recently launched Spectrum-Roentgen-Gamma mission \citep{Merloni12}. It is currently mapping the entire X-ray sky at 15'' spatial resolution, enabling the X-ray luminosity profiles of nearby ($z\approx 0.01$) $L^\star$ haloes to be spatially resolved. In a recent study, \citet{opp20b} used EAGLE and TNG to create mock {\it eROSITA} observations, and stacked these data about the co-ordinates of simulated star-forming and quiescent galaxy samples. This exercise demonstrated that if circumgalactic gas fractions are as sensitive to the properties of present-day $\sim L^\star$ galaxies as is indicated by the simulations, the differences will be evident in {\it eROSITA} data.

The correlation of galaxy properties with halo properties other than mass is often termed `galaxy assembly bias', an extension of the halo assembly bias that is characterised by the dependence of halo clustering (at fixed mass) on halo assembly time \citep[e.g.][]{sheth04,gao05}. The existence of galaxy assembly bias in cosmological hydrodynamical simulations was first demonstrated, also using the EAGLE simulations, by \citet[][see also \citealt{matthee17}]{chavesmontero16}, and has also been demonstrated in the IllustrisTNG simulations by \citet{monterodorta20}. Efforts to detect the effect in observational surveys have yielded mixed results \citep[see e.g.][]{yang06,deason13,tinker17,tojeiro17,zu17,wechsler18}, and at present the evidence remains inconclusive. A consequence of this bias is that it has the potential to introduce systematic errors into galaxy clustering diagnostics based on halo occupation distribution (HOD) models \citep[e.g.][]{peacocksmith2000}, since the latter assume that halo occupation is exclusively a function of halo mass \citep[e.g.][]{zentner14,zehavi18,artale18}. In the simulations presented here, galaxy assembly bias is manifest not only in the stellar mass content of dark matter haloes, but is also expressed in other properties of the galaxy-CGM ecosystem. Galaxy assembly bias may therefore also introduce a significant systematic error into the halo models of the cosmic atomic hydrogen distribution used to forecast the 21-cm emission power spectrum \citep[e.g.][]{padmanabhan17}. Such effects have already demonstrated in the \textsc{Shark} semi-analytical model \citep{chauhan20}, and are likely also present in cosmological hydrodynamical simulations. 

Finally, we remark that this galaxy assembly bias emerges as a consequence of physical interactions occurring throughout the growth of galaxies and their host dark matter haloes, mediated primarily by the growth of the central BH and complex gas dynamics governed by gravity and feedback-driven outflows. Since analytic techniques for connecting galaxies with dark matter structures, such as HOD and subhalo abundance matching, are typically stress-tested using semi-analytic galaxy formation models, an interesting avenue for future enquiry would  be a detailed comparison of the galaxy assembly bias signatures that emerge in those models with the signatures now becoming evident in cosmological hydrodynamical simulations of the galaxy population.

\section*{Acknowledgements}

We thank the referee for providing a helpful report, Ian McCarthy, Ben Oppenheimer and Joop Schaye for helpful discussions, and Diederik Kruijssen for advice concerning colour schemes suitable for colour blind readers. JJD acknowledges an STFC doctoral studentship and financial support from the Royal Society. RAC and AP are Royal Society University Research Fellows. This study has received funding from the European Union's Horizon 2020 research and innovation programme under grant agreement No. 818085 GMGalaxies, and was also partially supported by the UCL Cosmoparticle Initiative. The study made use of high performance computing facilities at Liverpool John Moores University, partly funded by the Royal Society and LJMU's Faculty of Engineering and Technology. The images in Figs. \ref{fig:DMimage} and \ref{fig:galaxypic} were produced using \texttt{py-sphviewer} \citep{sphviewer}.

\section*{Data Availability}

The data underlying this article will be shared on reasonable request to the corresponding author.



\bibliographystyle{mnras}
\bibliography{bibliography}





\bsp	
\label{lastpage}
\end{document}